\documentclass[aps,pra,amssymb,showpacs,supersriptaddress,floatfix,preprint]{revtex4-1}
\usepackage{amssymb}
\usepackage{amsmath}
\usepackage{graphicx}
\usepackage{bm}
\usepackage{caption}
\usepackage{subcaption}
\usepackage{bm}
\usepackage{color}
\usepackage[titletoc]{appendix}

\graphicspath{{images/}}
\begin{document}

\title{On the key role of oxygen vacancies electromigration in the memristive response of ferroelectric devices}

\author{C. Ferreyra$^{1,2}$, M. Rengifo$^{1,2}$, M.J. S\'anchez$^{1,3}$, A. S. Everhardt$^4$, B. Noheda$^4$, D. Rubi$^{1,2}$}


\affiliation{$^{1}$Instituto de Nanociencia y Nanotecnolog\'{\i}a (INN), CONICET-CNEA, Argentina \\
$^{2}$ Centro At\'omico Constituyentes, Av. Gral Paz 1499 (1650), San Mart\'{\i}n, Buenos Aires, Argentina\\
$^{3}$ Centro At\'omico Bariloche and Instituto Balseiro (Universidad Nacional de Cuyo), 8400 San Carlos de Bariloche, Río Negro, Argentina \\ $^{4}$ Zernike Institute for Advanced Materials, University of Groningen, 9747 AG Groningen, The
Netherlands  }
\date{\today}%


\begin{abstract}

Ferroelectric memristors are intensively studied due to their potential implementation in data storage and processing devices. In this work we show that the memristive behavior of metal/ferroelectric oxide/metal devices relies on the competition of two effects: the modulation of metal/ferroelectric interface barriers by the switchable ferroelectric polarization and the electromigration of oxygen vacancies, with the depolarizing field playing a fundamental role in the latter. We simulate our experimental results with a phenomenological model that includes both effects and we reproduce several non-trivial features of the electrical response, including resistance relaxations observed after external poling. Besides providing insight into the underlying physics of these complex devices, our work suggests that it is possible to combine non-volatile and volatile resistive changes in single ferroelectric memristors, an issue that could be useful for the development of neuromorphic devices.

\end{abstract}

\maketitle

\section{Introduction}

Memristors are defined as  capacitor-like structures displaying reversible and non-volatile electrical resistance changes upon the application of electrical stimulus \cite{saw_2008,iel_2016}. Besides their obvious potential for non-volatile memories, memristors can mimic the behavior of brain synapses and are therefore expected to constitute one of the building blocks of novel neuromorphic computing devices 
\cite{wan_2017,yu_2017} aiming to significantly enhance current electronic devices performance in complex tasks such as pattern recognition or big data analysis. 
Memristive effects usually rely in the electromigration of defects such as oxygen vacancies (OV), which can form conducting nanofilaments or modulate the height of Schottky barriers present at semiconductor/metal interfaces \cite{saw_2008}. 

Ferroelectric memristors \cite{chan_2012}, where the resistance change is linked to the switching of the ferroelectric polarization direction, present a high interest due to the electronic nature of this effect, which is significantly faster than ionic migration mechanisms. Therefore, faster processing capabilities are expected for these kind of devices. Ferroelectric memristive effects can be related to two mechanisms. On the one hand, for ultrathin ferroelectric layers (i.e. a few nanometers) sandwiched between asymmetric metallic electrodes, usually called ferroelectric tunnel junctions (FTJ), electronic tunneling effects occur and the barrier height depends on the orientation of the ferroelectric polarization \cite{ye_2009,tsy_2006,gar_2009,chan_2012}. The barrier height change $\Delta \Phi$ is given by $\Delta \Phi = ±P(\lambda_1-\lambda_2)q/\epsilon,$ where $P$ is the polarization, $\lambda_1$ and $\lambda_2$ are the electronic screening lengths for each  electrode, $q$ is the electronic charge and $\epsilon$ is the permitivity of the ferroelectric layer. The change  in the barrier height when the polarization is inversed produces a modification in the transmited current for a given applied voltage, and in this way  the resistance of the junction changes between two non-volatile states. We also notice that FTJ require a by-design asymmetry (i.e. different metallic electrodes) to give different resistive states for both polarization orientations. Interestingly, it was recently proposed that a FTJ-like behavior was found in a oxide heterostructure with no  ferroelectric layers, where a redox reaction between  MoSiO$_x$ and YBCO$_{1-x}$ layers was pointed as the dominant memristive mechanism for this system \cite{rou_2020}. These results highlight that memristive effects related to oxygen electromigration -possibly entangled with the electronic tunneling effect- should not be ruled out in the study of oxides-based FTJ. 

On the other hand, for devices involving thicker ferroelectric layers (i.e. tenths of nanometers) tunneling electronic transmission is negligible and the polarization-dependent memristive effect is usually attributed to the modulation of the Schottky barriers at ferroelectric/metal interfaces \cite{blom_1994, mey_2006, pin_2010}. A polarization pointing to the Schottky interface leads into a decrease of the barrier height, resulting in a low resistance state. When the polarization is inversed, the barrier height increases and a high resistance state is stabilized. Changes in interface energy barriers with polarization inversion have been inferred from spectroscopic techniques \cite{rau_2013, hub_2016}, transport measurements \cite{far_2014} and have also been theoretically simulated \cite{liu_2013}. A memristive effect could also be present in symmetric metal/ferroelectric/metal devices where two complementary interfaces can display memristive behavior simultaneously, as was previously shown for ionic migration-based memristive systems \cite{roz_2010}.

Different reports on ferroelectric memristors evidence a relaxation of the ideally non-volatile resistive states with time, with typical time-scale of minutes \cite{yin_2010,rub_2012}. This constitutes a severe drawback to the implementation of ferroelectric memristors in devices where retention times are a key issue. 

One possible scenario to explain the resistance relaxations in ferroelectic memristores is related to a loss of polarization with time due to the existence of a strong depolarizing field \cite{tia_2019}. On the other hand, it was shown that ferroelectric memristive effects coexist with OV  migration in different systems \cite{lu_2017, mao_2015, qia_2019, sul_2019, li_2015}. For instance, Qian et al. reported for BaTiO$_3$-based ferroelectric memristors that two  different  mechanism contribute  to  electroresistance (ER) \cite{qia_2019}. The knob that controls the appearance of one or other  mechanism was the time-length of the writing pulses; for short pulses (20 $\mu$s) the memristive behavior  was dominated by the modulation of the Schottky barrier by the direction of the ferroelectric polarization (that is, an electronic effect), but for longer pulses (20 s) OV electromigration seemed to dominate the change of the Schottky barrier height over the mentioned electronics effects. Both regimes  were clearly differentiated from remanent resistance vs. voltage curves with different circulations and distinct temperature behavior. 
For writing pulses with intermediate time-lenghts it is reasonable to expect the coexistence of both effects. 

These results indicate that further effort should be performed to disentangle the role of electronic effects and OV electromigration in the memristive behavior of systems including ferroelectric oxides. With this aim, in the present work we show for two different symmetric memristive metal/ferroelectric oxide/metal systems that distinct features of the   remanent resistance vs. voltage loops can be rationalized in terms of a competition between electronic effects, linked to the polarization inversion, and the electromigration of OV, both modulating the metal/ferroelectric Schottky barriers. 

Aided by phenomenological simulations that reproduce several non-trivial features of the experimental memristive reponse, we show that the driving force for OV electromigration in absence of external electrical stimulus is the depolarizing field present on the ferroelectric due to incomplete charge screening by the metallic electrodes. Moreover, our work provides strong evidence and physical insight about the coexistance of both non-volatile and volatile resistive changes in ferroelectric-based devices, an issue that might be useful for the development of neuromorphic computing devices.

\section{Experimental}\label{exp}

\begin{figure}[h!]
\centering
\includegraphics[scale=0.7]{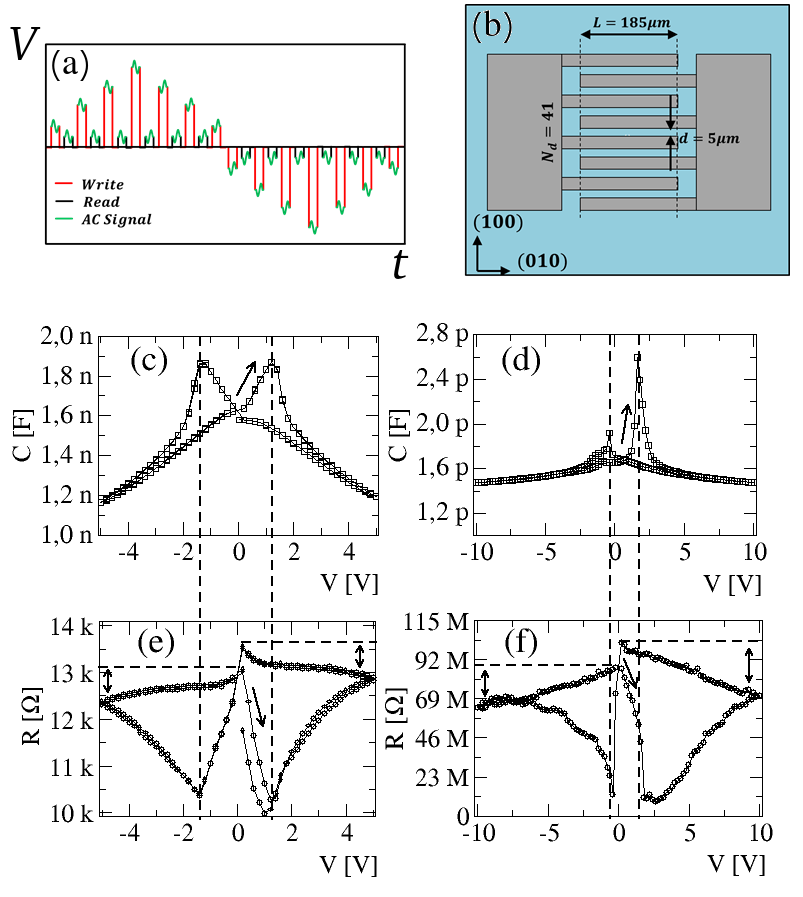}
\caption{(a) Sketch showing the stimuli protocol used to measure simultaneously C-V curves and HSLs in our samples. (b) Sketch displaying a top view of the interdigital Pt/SRO electrodes on top of BTO thin films. $N_d$ corresponds to the number of fingers of the electrodes. C-V curves measured on Pt/PZT/Pt (c) and SRO/BTO/SRO (d)  devices. HSLs recorded on Pt/PZT/Pt (e) and SRO/BTO/SRO (f) devices. See text for details. Single arrows indicate the evolution of the curves.}
\label{fig1}
\end{figure}

We have characterized two metal/ferroelectic/metal systems: a commercial Pt/PZT/Pt capacitor (Radiant Technologies AB ferroelectric capacitor) and a SrRuO$_3$/BaTiO$_3$/SrRuO$_3$ (SRO/BTO/SRO) device. In the first case, the polycrystalline PZT 20/80 layer is 255 nm thick, the Pt bottom and top electrodes are 150 and 100 nm thick, respectively, and the device area is $10^5\mu$m$^2$. In the second case, epitaxial BTO thin films (90 nm thick) were deposited by pulsed laser deposition on NdScO$_3$ substrates. After that, interdigital SRO/Pt top electrodes were deposited and shaped by standard lithographic techniques. The device geometry is shown in Figure \ref{fig1}(b). Further exhaustive characterization of the BTO-based structures can be found in references \cite{ever_2016,ever_2019,ever_20}. Electrical measurements were performed along the BTO (100) in-plane direction.

We notice that we chose to perform electrical AC measurements due to the difficulty in performing standard DC characterization given the very high 2-points resistance of both systems ($>$10$^8$ $\Omega$). We recorded simultaneously capacitance-voltage (C-V) and remanent resistance-voltage (usually called Hysteresis Switching Loops or HSLs) curves from AC measurements by using a standard LCR-meter set in RC parallel circuit mode, which allows extracting values for R and C. 

The stimuli protocol consists on the application of alternated writing and (undisturbing) reading DC pulses, with a superimposed small AC signal in both cases (frecuencies $f$ between 100Hz and 1MHz were tested), as shown in Fig.\ref{fig1}(a). The time-width of writing pulses was $\tau_{W} \approx$ 50ms and the time between two consecutive writing pulses was $\Delta\tau$ $\approx$ 4s. The C-V curve is extracted from the capacitances measured during the application of the writing pulses, while the HSL is extracted from the resistances measured during the application of reading pulses. The DC bias of the reading pulses was 200mV and the AC amplitude -tipically a few hundred mV- was chosen at each frequency to maximize the signal-to-noise ratio of the HSL. The C-V curves ($f$=100kHz) for Pt/PZT/Pt and SRO/BTO/SRO devices are shown in Figs.\ref{fig1}(c) and (e), respectively, displaying in both cases the typical butterfly-like shape of ferroelectric systems. A slight imprint is found in the case of the SRO/BTO/SRO device.

Figures \ref{fig1}(d) and \ref{fig1}(f) display HLS's recorded for Pt/PZT/Pt and SRO/BTO/SRO devices, respectively, also for $f$=100kHz. In both cases, it is found the so-called “table with legs” \cite{roz_2010}, typical of two series complementary memristive interfaces; that is, when one interface switches from low (LR) to high resistance (HR), the other switches inversely. Two series switchable Schottky interfaces (Pt/PZT and  SRO/BTO, respectively \cite{scott_99,chen_09,liu_13}) are likely at the origin of the memristive behavior in each system. We notice that in both cases, the “legs” in Figures \ref{fig1}(d) and \ref{fig1}(f) match well with the capacitance peaks related to ferroelectric switching  observed, respectively, in Figs. \ref{fig1}(c) and \ref{fig1}(d). This suggests the existence of ferroelectric memristive behavior; that is, each interface’s Schottky barrier is  modulated by the direction of the ferroelectric polarization. 

However, a distinct feature is observed in both HSLs: the resistance of the top part of the ``tables” presents a noticeable positive (negative) slope for negative (positive) voltages. This behavior is at odd with the one exhibited by (non-ferroelectric) symmetric  memristive complementary interfaces \cite{roz_2010, fer_2020}, where the ``table with legs” HSLs exhibit a flat ``table" (that is zero slope for both positive and negative voltages), as it is sketched by the horizontal dotted lines in Figures \ref{fig1}(e) and (f). This quite unusual feature  suggests that each interface resistance is modulated by the competition of ferroelectric memristive effects \textit{and}  the electromigration of OV. A similar interpretation was done by Hu et al. from the observation of a non-squared HSL in BiFeO$_3$-based FTJ \cite{hu_2016}. Going back to our experiments, we recall that HSL for the Pt/PZT/Pt system were recorded in a broad frequency range between 100Hz and 1MHz and displayed similar features as those already described  \cite{SI, nov07}

In order to get further insight into the different scenarios and explain our experimental HLSs, we have performed numerical simulations based on the voltage enhanced OV drift (VEOV) model, developed originally to describe 
 non-ferroelectric perovskites \cite{roz_04, roz_2010, rub_2013}, further extended to  simple oxides based \cite{ghe_2013, fer_2020}
 memristive devices and here adapted  to a symmetric metal/ferroelectric/metal system, as it will be shown below.

\section{SIMULATIONS}

To simulate the memristive behavior of the devices, we  consider a symmetric metal/ferroelectric/metal (M/FE/M) system where both metal/ferroelectric interfaces are Schottky type and connected in a back-to back configuration.
The barrier heights depend on the direction of the electrical polarization and on the local concentration of OV at the interfaces.

We assume for the ferroelectric oxide that the presence of OV locally reduce its resistivity, as is the case of BTO \cite{yang_04}, PZT \cite{chentir_08} and other simple oxides such as TiO$_2$ \cite{ghe_2013}. 

Bearing these considerations in mind, we propose for the two-point resistance of the device the following phenomenological equation:

\begin{equation}  \label{e1}
 R^{T}=R^{eff}A^{FE}(P)M^{OV},
\end{equation}

being $R^{eff}$ an effective (constant) resistance, while   $A^{FE}(P)$ and  $M^{OV}$ are two adimensional factors. $A^{FE}(P)$ modulates the interfaces resistances and  is a function of the actual ferroelectric polarization $\vec{P}$, while  $M^{OV}$ depends on  the OV distribution along the complete device, as we will show below.

We start by deriving  the relation between $A^{FE}$ and the device polarization $\vec P$ (which we assume as single domain) by considering, in the first place, a scenario with no OV electromigration.  
As it is customary  in memristive devices, the HSL is constructed by applying a read pulse $V^{R}$ after the write one $V^{W}$, which induces the resistive change. $V^{R}$ is small enough to warrant that both Schottky interfaces, named L and R respectively, remain reversely biased. The current $I$ for a Schottky interface biased with a voltage $V$ can be written as \cite{sze}:

\begin{equation}\label{I}
I=\beta(V,T) \exp{[-\frac{e}{k_BT}\phi_{\alpha}}],
 \end{equation}
being $\beta (V,T)$ a constant  for  given temperature T and voltage bias $V$.
The  barrier height $\phi_{\alpha}$ 
for  a metal/ferroelectric Schottky junction can be expressed as \cite{hub_2016,sten_2011}

\begin{equation}\label{phi}
\phi_\alpha=\phi_\alpha^0 \mp \gamma |\vec{P}|,
 \end{equation}

where $\phi^{0} _{\alpha}$ is the  bare barrier height -related to the difference between the metal work function and the insulator electron affinity- 
and $\mp \gamma |\vec{P}|$ is the correction 
due to the presence of ferroelectric bound charges at the interface. If the polarization points to the Schottky interface then the barrier height will decrease, while in the opposite case the barrier height will increase.

\begin{figure}[h!]
\centering
\includegraphics[scale=0.5]{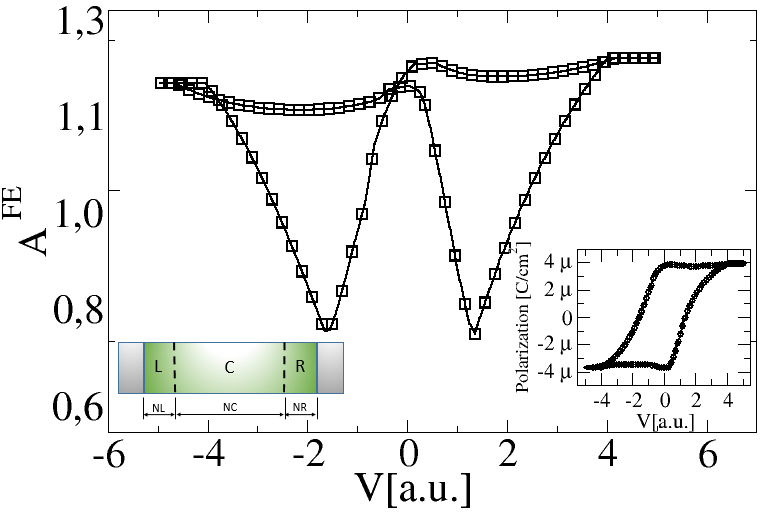}
\caption{$A^{FE}$ vs. $V^{W}$, being ${A}^{FE}$  defined in Eq.\eqref{Afe} (the value of $\tilde{\gamma}$ is given in Table 1). Left inset: sketch of the metal/ferroelectric/metal device, where L, C and R zones defined for our simulations are explicitely shown. NL, NC and NR correspond to the number of nanodomains present in each zone, respectively. Right inset: $P-V$ loop obtained after integration of the experimental $C-V$ data for the  Pt/PZT/Pt system. Notice that the voltage scale of the $P-V$ loop was converted according to 1 V = 1 a.u. in order to match the voltage scale of the simulation showed in the main panel}
\label{fig2}
\end{figure}

We now consider the symmetric M/FE/M system comprising the two series back-to-back Schottky interfaces L and R, as it is sketched in the inset of Figure  \ref{fig2}.  We assume for the polarization $P$ a positive value when  pointing from L to R, and a negative value otherwise. Thus, when  $P$ goes from L to R, the  barrier height  at interface L will increase while  at interface R will  reduce.
In this case,  the current $I$ will be limited by the  barrier L,  where most of the voltage drop due to $V^{R}$ occurs, and according to the previous convention for the polarization sign, it can be expessed  as   
$I\sim I_L =\beta \exp{[-\frac{e}{k_BT}(\phi_{\alpha}^0+\gamma_L P)}]$. 

On the other hand, when the polarization is reversed ($P<0$), the current will be limited by interface R and thus given by 
$I\sim I_R=\beta \exp{[-\frac{e}{k_BT}(\phi_{\alpha}^0-\gamma_R P)}]$. 

The last two equations can be written in compact form as:

\begin{equation}\label{I3}
I=\beta\exp{[-\frac{e}{k_BT}(\phi_{\alpha}^{0}+{\gamma} |P|)]=I_0\exp(-\tilde{\gamma}|P|)},
\end{equation}
where we assume that both interfaces are identical
($\gamma_L= \gamma_R= \gamma$) and we define the constants $I_0=\beta\exp{[-\frac{e}{k_BT}(\phi_{\alpha}^{0})}]$ and  
$\tilde{\gamma}= \frac{e}{k_BT} \gamma $, respectively. 

Equation \eqref{I3} explicitely shows  the dependence of the current  $I$ along the device with the polarization $P$. 

Under the assumption  that the device resistance is dominated by the interfaces and solely modulated by the direction of the ferroelectric polarization, a  remanent  two point resistance $R(P)$ can be computed as: 

\begin{equation}\label{Rrem}
R(P)\approx\frac{V^{R}}{ I_0\exp(-\tilde{\gamma}|P|)}=\frac{R_0^{}}{\exp(-\tilde{\gamma}|P|) },
\end{equation}
being  $R_0^{}= V^{R}/ I_{0}$, with resistance dimension. By comparing Eqs. \eqref{e1} and \eqref{Rrem}, and for the scenario with no OV electromigration considered so far, we assume   $R^T\approx R(P)$  and thus  $R_0 = R^{eff}M_0^{OV}$, being $M_0^{OV}$  a constant determined by the initial OV configuration. Therefore  we obtain :

\begin{equation}\label{Afe}
A^{FE}(P)\equiv \frac{1}{\exp(-\tilde{\gamma}|P|)},
\end{equation}
which gives the modulation of the device resistance with the amplitude and direction  of the ferroelectric polarization $\vec{P}$.

To compute the evolution of $A^{FE}$ with $P$ upon voltage cycling, it is necessary to know the dependance of $P$ on V$^W$. To this end, we use the $P-V$ loop corresponding to the  Pt/PZT/Pt device, obtained after integrating the experimental $C-V$ data (Figure \ref{fig1}(c)), which shows a more symmetric behavior than the $C-V$ curve corresponding to the SRO/BTO/SRO device (Figure \ref{fig1}(d)).

The $P-V$ loop is shown in the (right) inset of Fig. \ref{fig2}, while in the main panel is displayed the calculated $A^{FE}$ vs. $V^{W}$. 
The latter figure can be interpreted (up to dimensional prefactors) as 
a HSL obtained under the assumption  that the memristive effect relies exclusively on the modulation of both Schottky barriers by the switchable ferroelectric polarization. The  curve is qualitatively similar to standard  HSLs reported for symmetric memristive systems \cite{chen_2005,roz_2010, fer_2020} in which the resistance switching is
based on a completely different mechanism involving OV electromigation. 

In our case, the obtained $A^{FE}$ vs $V^{W}$  is a demonstration of the complementary behavior in M/FE interfaces.
However, it significantly  differs  from  the  experimental  HSL  displayed  in  Figure  \ref{fig1}(e), mainly in the slopes of the top zones of the ``table", as already described in the Sec.\ref{exp}. This indicates the need of including an additional ingredient to properly simulate the behavior of our device. 

In the following, we will show that by taking into account the electromigration of OV in the modelling as a competitive memristive effect with the ferroelectric one, allows to properly collect the physics of our devices.


To determine the $M^{OV}$ factor in Eq.\eqref{e1}, related to OV electromigration, we consider an unidimensional path across the ferroelectric which bridges both electrodes and is divided in the three zones: the central one C, composed of NC nanodomains, and the left (L) and right (R) zones, having NL and NR nanodomains, respectively, comprising the interfaces with the metallic electrodes. The total number of nanodomains is thus given by N= NL+NC+NR. 

To each  nanodomain we assign a site  $i$ and define its resistivity,  $\rho_{i}$,  in terms of  the local density of OV, $\delta_{i}$.
According to \cite{ghe_2013,fer_2020} we write: 

\begin{equation}  \label{e2}
 \rho_{i}^{OV}=  \rho_{0} (1 - A_{i} \delta_{i}),
\end{equation}

where $\rho_{0}$ is the  resistivity in absence of OV.
The factors  $A_i$ satisfy $ A_{i} \delta_{i} < 1 \; \forall i$, and for simplicity are taken as constants along each specific region L, C and R, respectively.
In the simulations we consider A$_L$ $\sim$ A$_R$ $<$ A$_C$, to capture  the fact that the  regions close to the Schottky barriers  are  more resistive that the central  C zone.

For a given OV configuration, the two point  resistance $R^{OV}$ of the complete domains path is 
\begin{equation}  \label{Rrs}
 R^{OV} =R_{0}' (N - \sum_{i=1}^{NL}A_{L} \delta_{i}- \sum_{i=NL+1}^{NL+NC}A_{C} \delta_{i}-\sum_{i=N-NR+1}^{N}A_{R} \delta_{i}),
\end{equation}

where we have defined  the resistance $R_{0}'= \rho_{0}{\cal L}$, being $\cal L$ the typical lenght of each nanodomain.

We  now define the adimensional prefactor  $M^{OV}$  appearing in Eq.\eqref{e1} as: 
\begin{equation}  \label{Mrs}
 M^{OV} \equiv \frac{ R^{OV}} {R_{0}'}
\end{equation}
which is computed straigthforwardly from Eq.(\ref{Rrs}), once $R^{OV}$ is known. 

In the VEOV model \cite{roz_2010}, the probablity rate $p_{ij}$ for an OV to hop from one site to a neighbour site is be given by

\begin{equation}\label{proba}
p_{ij}=\delta_i(1-\delta_j)\exp(-V_{0}+\Delta V_i),
\end{equation}

where $\delta_i$ and $\delta_j$ are the  OV densities at sites $i$ and $j$ respectively.
We define $V_{0}$ as the activation energy for OV diffusion -we take as $V_{0L},V_{0C}$ or $V_{0R}$ depending on the region L,C or R, respectively- and   $\Delta V_i$ as the voltage drop at site $i$.  Both are   adimensional quantities written in terms of the thermal energy $k_B T$.

In the following we assume  that $\Delta V_i$  can be written as the sum of two terms. The first one is the voltage drop due  to the external bias $V^{W}$ and the second  is the contribution from a depolarizing voltage arising from the existance of a depolarizing field $E_{dp}$ due to incomplete screening of the ferroelectric bound charges at both M/FE interfaces \cite{desu_1995}. 

We assume for the depolarizing field $E_{dp}= -\beta P$ \cite{dawber_2003}, being $P$ the ferroelectric polarization and $\beta$ a constant.

Therefore, the voltage drop $\Delta V_i$ at site $i$ is written as

\begin{equation}\label{deltav}
\Delta V_i=V^{W}  \frac{R^{OV}_i}{R^{OV}}   + E_{dp}{\cal L} = V^{W} \frac{R_i^{0V}}{R^{OV}} - \xi P,
\end{equation}


where $R^{OV}_i= \rho^{OV}_i {\cal L}$ and $\xi= \beta {\cal L} $.

\begin{figure}[h!]
\centering
\includegraphics[scale=0.5]{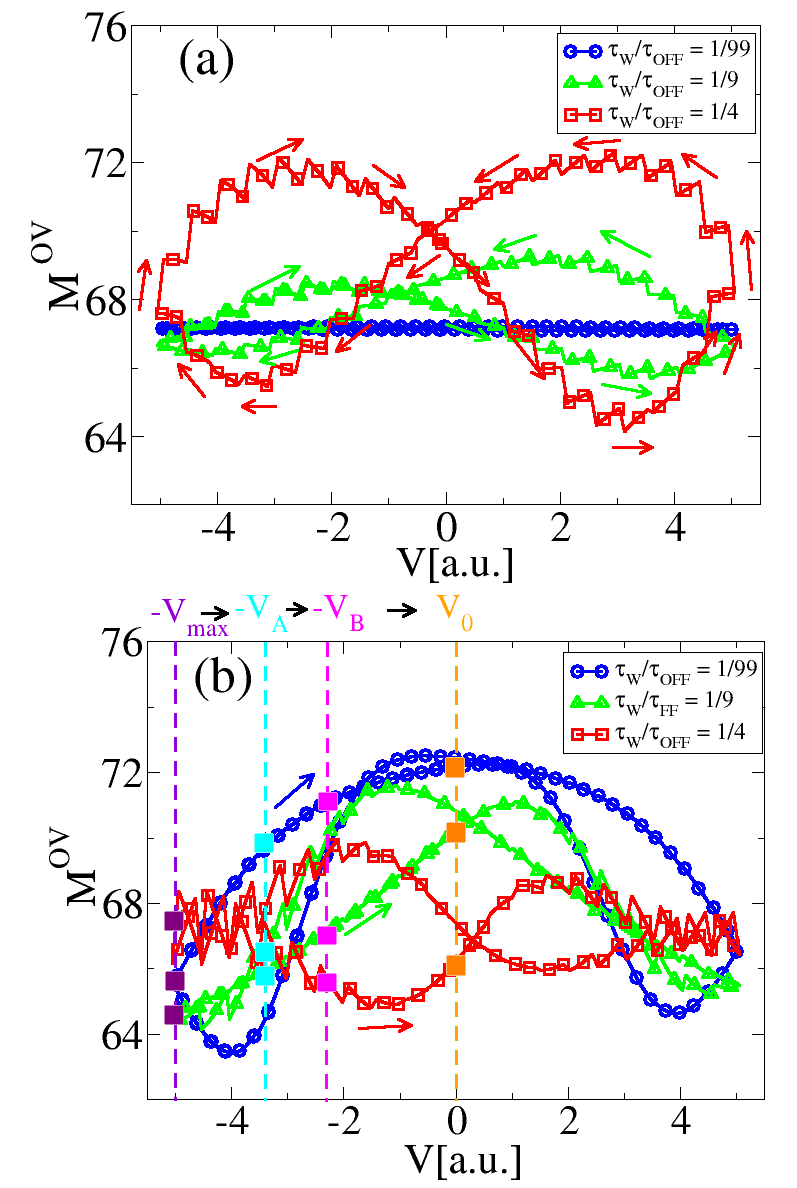}
\caption{Evolution of $M^{OV}$ vs. $V^{W}$ for different $\tau_W$/$\tau_{OFF}$ ratios, shown in the respective legends. 
(a) Under the sole action of the external stimuli (voltage). (b) Including the effects of both the external stimulus and the depolarizing field.}
\label{fig3}
\end{figure}


Given an external  voltage $V^{W}(t)$ applied  at time $t$, the OV density at site \textit{i}  is updated for each simulation step  according to Eqs. \eqref{proba} and \eqref{deltav}. 

In order to restrict the dynamics of OV to the active region,  we take  the rates $p_{01} =p_{10}=p_{N N+1}=p_{N+1 N}=0$. In addition, as the total number  of OV  is conserved, it is satisfied  that $\sum_{i=1}^{N} \delta_{i}= 1 \,, \forall t$.

To match the experiments, we chose the  stimulus  $V^{W}(t)$ as  a linear ramp  following the cycle $0 \rightarrow V^{W}_{m1} \rightarrow -V^{W}_{m2} \rightarrow 0$.
At each simulation time step $t_k$ we compute the local voltage drops ${\Delta V}_i (t_k)$ and   employing the probability rates $p_{ij}$ we obtain the transfers between nearest neighboring sites. 
Afterwards the values $\delta_i(t_k)$ are updated to a new set of densities $\delta_i(t_{k+1})$,
with which we compute, at time $t_{k+1}$, the local resistivities  $\rho_i(t_{k+1})$, the local voltage drops under the applied voltage $V^{W}(t_{k+1})$ and the factor $M^{OV}(t_{k+1})$ from Eq.(\ref{Mrs}), to start the next simulation step at $t_{k+1}$. For the polarization $P(V^W(t_k))\equiv P(t_k)$ we use the Pt/PZT/Pt experimental curve given by Figure \ref{fig2}.


Following this alghoritm, we compute the evolution of the $M^{OV}$ vs $V^W$, 
which accounts in Eq. \eqref{e1}  for  the memristive effect related to OV electromigration due to the action of both the external voltage and the internal depolarizing field. 

We recall that in standard memristive systems driven by OV electromigration the time-widths and intensity of writing pulses are key parameters that tune the memristive response \cite{acevedo_2018}. For the present ferroelectric system, the dynamics of OV will be dependant on two competitive effects. When the write voltage $V^W$ is ON, OV  drift in one direction (i.e. parallel to the applied external electrical field, which also aligns the polarization in the same direction). When $V^W$ is turned OFF, OV electromigration  is  solely  ruled by the depolarizing field, and thus in the direction anti-parallel to the polarization.

It is therefore critical for the overall OV dynamics the time windows in which the external voltage is switched ON and OFF, respectively. To collect these features in our simulations, we simulated three situations with $\tau_W$/$\tau_{OFF}$ = 1/99, 1/9 and 1/4, where $\tau_W$ and $\tau_{OFF}$ are the ON and OFF external voltage time windows, respectively. We assume in all cases that two consecutives write pulses are separated by a time-width of 1000 a.u. 

In Figure \ref{fig3} (a) we show the evolution of $M^{OV}$ as a function of $V^{W}$, without including the effect of the depolarizing field (i.e. perfect screening at the electrodes) for these three situations. The numerical values used for the simulation parameters are shown in Table \ref{tab}. It is found that for $\tau_W$/$\tau_{OFF}$ = 1/4 there is a large hysteresis in the curve, for $\tau_W$/$\tau_{OFF}$ = 1/9 the hysteresis is mild while no hysteresis is found for $\tau_W$/$\tau_{OFF}$ = 1/99. This indicates that, as expected, longer external stimulus induces larger OV electromigration in the direction parallel to the external stimulus, similar to the case of non-ferroelectric systems. 

\begin{figure}[h!]
\centering
\includegraphics[scale=0.7]{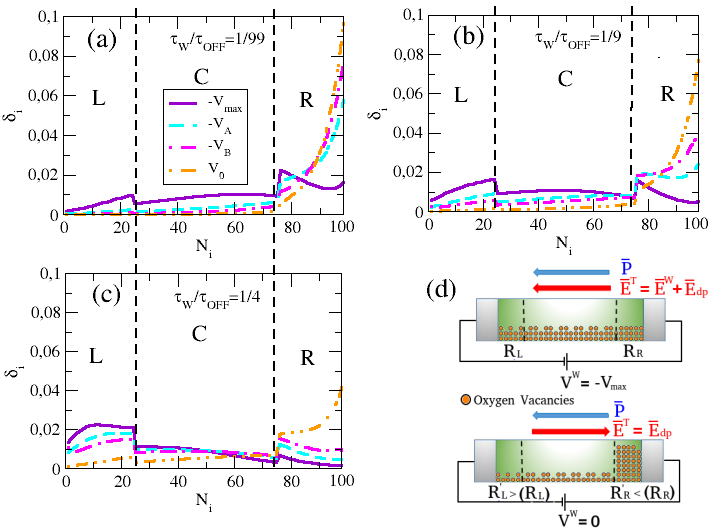}
\caption{OV profiles for $\tau_W$/$\tau_{OFF}=$ 1/99(a) 1/9 (b) 1/4 (c). The profiles with different colours follow the application of the writing voltages displayed in Fig. \ref{fig3}(b) (same colour codes). (d) Sketch displaying OV profile upon the application of $V^W=-V_{MAX}$ (upper panel) 
and its evolution afterwards upon $V^W=V_0=0$ (lower panel). The electrical field $\bar{E}_T$ is the superposition of the external field $\bar{E}^W$ (pointing to the left in the upper panel and zero in the lower one) and the depolarizing field $\bar{E}_{DP}$ (pointing to the right in both cases). OV migration from left (L) to right(R) is driven by $\bar{E}_{DP}$. The device resistance increase is dominated by the decrease of OV (increase of oxygen content) at the L interface.}
\label{fig4}
\end{figure}

Figure \ref{fig3}(b) analyzes a similar situation now including the presence of the depolarizing field $E_{dp}$ (again, the parameters are shown in Table \ref{tab}). In this case, unlike the previous one, there is an evident large hysteresis for the three situations. We notice that for $\tau_W$/$\tau_{OFF}$ = 1/99 and 1/9 the curves circulations are inverted with respect to the one corresponding to 1/4. This crossover is an indication of a change in the OV dynamics between shorter and longer pulses, where the effect  of the depolarizing field should become more pronounced as $\tau_W$ is reduced. 
In order to shed light on this issue, Figs. \ref{fig4}(a)-(c) displays the OV profiles for the explored $\tau_W$/$\tau_{OFF}$ ratios, after the application of writing pulses $-V_{MAX}\rightarrow -V_{A}\rightarrow -V_{B}\rightarrow V_{0}=0$, as indicated in Fig. \ref{fig3}(b). Notice that the colour code of the OV profiles of Figs.\ref{fig4}(a)-(c) corresponds to the $V_{i}$ colours in Fig.\ref{fig3}(b). 

The analysis of Figs. \ref{fig4}(a)-(c) shows that for all $\tau_W$/$\tau_{OFF}$'s, the OV profiles after $-V_{MAX}$ are distributed in the three zones of the device (L,C and R) with no presence of sharp peaks. In this scenario, both interfaces L and R contribute to $M^{OV}$. Interestingly, after $V^W= V_{0}=0$ most OV migrate to the R interface, and the driving force for this movement is the depolarizing field pointing to the R interface  (we recall that $P$ points to the left inteface L after $-V_{MAX}$). For $\tau_W$/$\tau_{OFF}$ = 1/99 and 1/9 this electromigration process is fast, reflected in a monotonous increase of $M^{OV}$ as the writing pulses go from $-V_{MAX}$ to $V_{0}$. 
The final OV configuration suggests that the main contribution to $M^{OV}$ comes from the L interface while the R interface -with a high OV density- presents a marginal contribution.

In the case of $\tau_W$/$\tau_{OFF}$ = 1/4 (longer writing pulses), the overall behaviour is the same (transfer of OV from L to R interfaces) but with a slower dynamics, which produces initially  a slighty erratic curve followed by  a non-monotonic evolution of $M^{OV}$ when the writing voltage goes from $-V_{MAX}$ to $-V_{0}$. Notice that between $-V_{MAX}$ and $-V_{B}$, $M^{OV}$ decreases and further increases between $-V_{B}$ to $V_{0}$. This behavior reflects the competition between the L and R interfaces to the value of $M^{OV}$, the first (second) one increasing (decreasing) its contribution to the resistance as it becomes drained (filled) of OV. Therefore, wider writing pulses partially delay (and thus counterback) the action of the depolarizing field, unveiling the described (more complex) OV dynamics.

\begin{figure}[h!]
\centering
\includegraphics[scale=0.5]{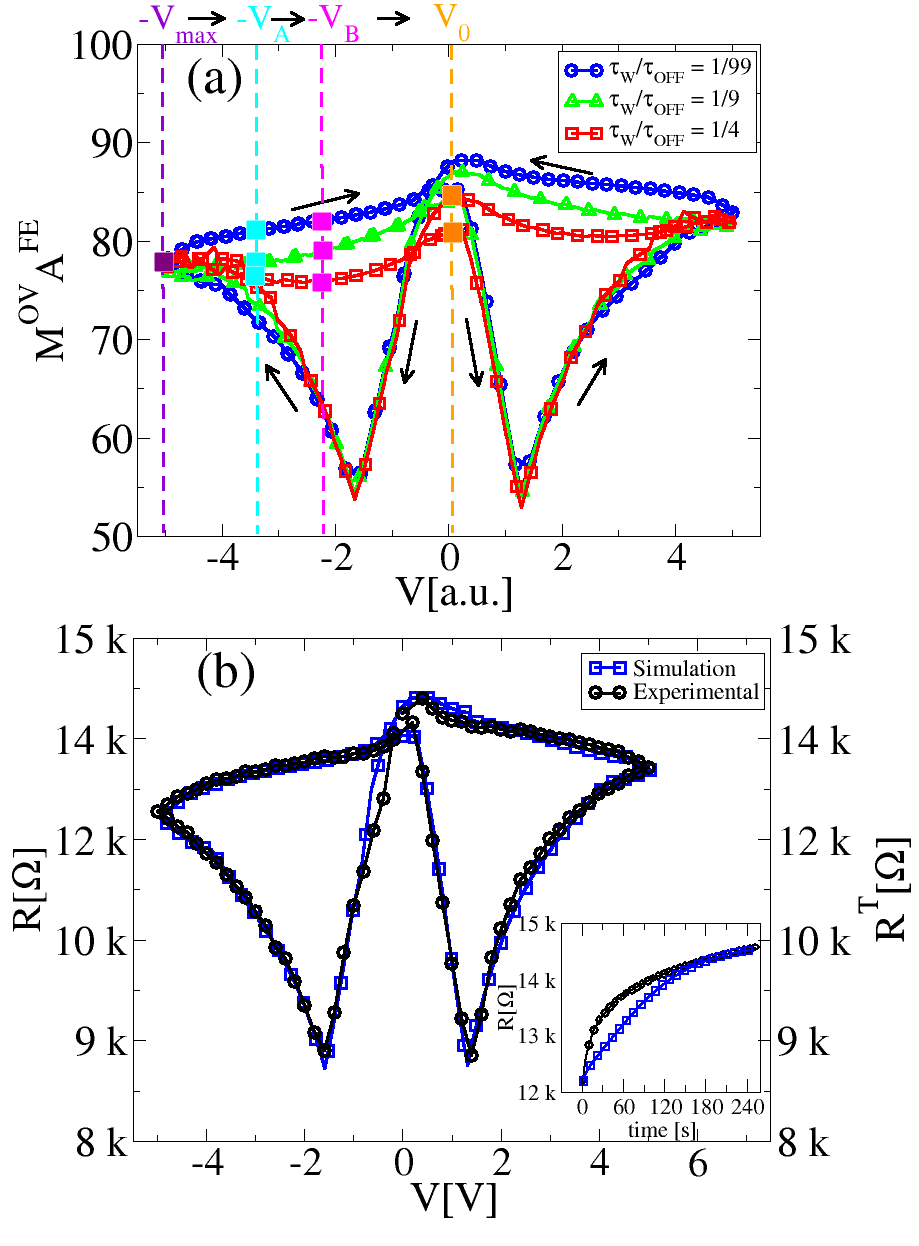}
\caption{(a) Evolution of $M^{OV} A^{FE}$ vs. $V^{W}$ for different $\tau_W$/$\tau_{OFF}$ ratios.  (b) simulated  HSL (right vertical axis) and experimental HSL (left vertical axis) corresponding to  the Pt/PZT/Pt device (Fig. \ref{fig1}(e)). In (a) and (b)  the simulations include the effect of OV electromigation due to both the external voltage and the depolarizing field together with the modulation of metal/ferroelectric barriers by the ferroelectric polarization.  Inset:  Experimental (dark)  and simulated (blue)  relaxation of the resistance with time after a single write pulse $V^W$ followed by reading pulses. The numerical values of  parameters are given in Table \ref{tab}. The conversion factor for $V^W$  between (a) and (b) is  1 a.u. = 1 V.}
\label{fig5}
\end{figure}

Finally, Fig. \ref{fig5}(a) displays the evolution of the product  $A^{FE}M^{OV}$ as a function of $V^{W}$. 
In this case, both the ferroelectric and OV electromigration (including the effect of the depolarizing field) 
memristive effects are included. 

We notice that the maximum depolarizing field reached in the simulations was lower than the coercive field 
$E_C$ (roughly $E_{DP} \leq$ 0.5 $E_C$), in order to describe a realistic situation where no strong sample depolarization takes place. 

For $\tau_W$/$\tau_{OFF}$ = 1/4 (long writing pulses) the simulated $M^{OV} M^{FE}$ displayed in Fig. \ref{fig5}(a) evidences that the remanent resistance values after $-V_{MAX}$ and $-V_{0}$ are  similar. In other words, the HSL exhibits 
a flat top-part of the ``table". For shorter writing pulses -$\tau_W$/$\tau_{OFF}$ = 1/9 and 1/99- where the effect of the depolarizing field on OV dynamics becomes largely dominant, we find that the remanent resistance values after $-V_{MAX}$ is smaller than after $V_{0}$. This shows that the top-part of the HSLs present non-zero slopes mimicking the experimental HSLs displayed in Figs. \ref{fig1}(e) and (f), respectively. 
This indicates that for our experimental systems, in the range of used writing stimulus (voltage and pulse time-widhts), OV dynamics is mainly controlled by the depolarizing field. \\
With the evolution of $A^{FE}M^{OV}$ vs. $V^W$ already figured out, according to Eq.(\ref{e1}) we can fit the experimental HSL by determining the multiplicative constant $R^{eff}$, which will depend on the initial OV profile. We recall that as the experimental $R$ is an effective value dependant on the reading frequency $f$ (see Section II and the Suppl. Info.), $R^{eff}$ should be also re-scaled properly for each experimental reading frequency. Figure \ref{fig5}(b) displays the 100kHz experimentl HSL for the Pt/PZT/Pt system and the corresponding simulation. 
For the simulation we use $\tau_W$/$\tau_{OFF}$ = 1/99, which is very close to the experimental ratio  $(\tau_W$/$\tau_{OFF})_{EXP}$ $\approx$ 0.012 given by the used stimulus protocol (see Sec.\ref{exp}  for details).  For the voltage scale, we considered 1 a.u. = 1 V.  The simulated curve displays an excelent agreement with the experimental data, with all the relevant features of the experimental HSL  collected by the simulated one. We also notice that we constrained the quantitative simulation to the Pt/PZT/Pt system as the SRO/BTO/SRO device presents a slight asymmetric electrical response (see Figure \ref{fig1}(f)), reflecting some asymmetry between both metal/oxide interfaces. A quantitative modelling of this system requires to reflect this issue in the initial assumptions of the simulation and goes beyond the scope of the present paper.


\begin{table}[h!]
\caption {Table with the numeric values of the different parameters used in the simulations. The equations where the parameters have been used are numbered as in the main text and shown in the right column.}
\label{tab} 
\begin{tabular}{|l|l|l|l|}
\hline
\multicolumn{1}{|c|}{\textbf{Parameters}}        & \multicolumn{1}{c|}{\textbf{Value}} & \multicolumn{1}{c|}{\textbf{Equation}}                                                     \\ 
\hline
${\cal L}$                                & 2.55 nm                          & (8)                                                                                \\
\hline
NL                                    & 61.20 nm                          &(8)                                                                                 \\

\hline
NC                                     & 132.60 nm                          & (8)                                                                                                                         \\ 
\hline
NR                                        & 61.20 nm                          &(8)                                                                                                          \\ \hline
$\tilde{\gamma}$                                  & $1.06.10^5$ cm$^2$/C                                                                         &(4)                                                                                          \\ \hline

$R^{eff}$                                                   & 160 $\Omega$                                                                      & (1) 
                                         \\ \hline
$R_0'$                                                    & 129 $\Omega$                       &(8)                                                                                                                                             \\ \hline
$R_0$                                                 & 11.60 k$\Omega$                                                           &(5)                                                                                   
                               \\ \hline

$A_{L}$                                        & 10.40                                                           & (8)                                                                           \\ \hline
$A_{C}$                                        & 26.40                                                         & (8)                                                                          \\ \hline
$A_{R}$                                        & 11.04                                                        & (8)                                                                             \\ \hline
$V_{0L}$                                          &2.9* (0.07 eV)                                                                                           & (10)                                                                              \\ \hline
$V_{0C}$                                          & 2.3* (0.06 eV)                                                                                                                                        &(10)                                   \\ \hline
$V_{0R}$                                          & 3.2* (0.08 eV)                                                                                              & (10)                                                                         \\ \hline

$\xi$                                        & $2500$ Jcm$^2$/C$^2$                                             & (11)                                                                               \\ \hline
\end{tabular}
    The parameters $V_{0L}$, $V_{0C}$ and $V_{0R}$ are normalized in units $K_B T= 0.025$ eV (room temperature).

\end{table}


Finally, we notice that in the scenario discussed above, where OV drift  is strongly driven by the depolarizing field acting in absence of external stimuli, remanent resistance states should display time relaxations after the application of single write pulses. In order to explore this, we have  performed an additional experiment in our Pt/PZT/Pt systems. A voltage writing pulse (+6V, larger that the maximun voltage excursions of the HSL showed in Figure \ref{fig1}(e)) was applied, followed by reading the effective remanent resistive states as a function of time (a small undirsturbing DC pulse, with a superimposed AC signal of 100kHz, was used to extract the remanent resistance values with the LCR meter). 

The inset of  Fig.\ref{fig5} (b) displays that, as expected, the remanent resistance relax with time, with typical time-scale of minutes \cite{com}, reflecting the action of the depolarizing field on OV electromigration. 

Upon the application of the +6V pulse the polarization points from right (R) to  left (L), giving an OV profile which is the mirror of the one  displayed in Figures \ref{fig4}(a) or (b). After the writing pulse is switched off, OV start to migrate from R to L and thus the device resistance starts to be dominated by and increases at the R interface. The experimental relaxations were simulated using the model described above, as also shown in the inset of  Fig.\ref{fig5}. It is found a very good agreement between the experimental and simulated curves for $t \geq 120s$ after the pulse application. For shorter times, the experimental relaxation does not follow a pure exponential behavior -as it is the case of the simulated curve- and this produces a slight divergence between both curves. This can be related to the existance of effects not included in our modelling, such as the presence of charge trapping/detrapping process at the interfaces that could affect the device electrical transport and resistive memory behavior \cite{deli_2006,shang_2006}. A quite similar resistance evolution (initial drop and further increase with time, not shown here) was found upon and after the application of negative single pulses, as expected given the device and the corresponding HSL symmetry.


\section{Conclusions}

In summary we have found that, besides the modulation of the interfacial barriers by the direction of the ferroelectric polarization, the electromigration of OV 
appears  as a  competing effect  in the memristive response of symmetric metal/ferroelectric/metal devices.
The driving force for OV electromigration is the depolarizing field linked to incomplete charge screening at ferroelectic/metal interfaces. With these ingredients, we have been able to model  and reproduce several non-trivial features of the experimental electrical behavior, including HSL and resistance relaxations. Our work sheds light on the non trivial interplay between  two relevant  mechanisms that contribute to the resistance change in ferroelectric memristors, and in particular highlights the key role played by OV electromigration. 
We recall that the combination of both non-volatile and volatile (relaxations) resistance changes offers oportunities for the development of devices with neuromorphic capabilites, able to mimic the  behavior of synapses and neurons, respectivey, in a single device. Moreover, the volatile effect could be tailored by engineering the interfaces in order to control the screening lenghts and tune the intensity of the depolarizing field.

\section*{Acknowlegments}
 
We thank Dr. S. Matzen for the microfabrication of BTO devices and Dr. L. Steren for providing the PZT capacitor. We acknowledge support from INN-CNEA, UNCuyo (06/C591), ANPCyT (PICT2014-1382, PICT2016-0867, PICT2017-1836) and EU-H2020-RISE project "MELON" (SEP-2106565560). DR thanks the Nederlandse Organisatie voor Wetenschappelijk Onderzoek for a visiting grant.


 
\bibliography{references}

\begin{thebibliography}{49}%
\makeatletter
\providecommand \@ifxundefined [1]{%
 \@ifx{#1\undefined}
}%
\providecommand \@ifnum [1]{%
 \ifnum #1\expandafter \@firstoftwo
 \else \expandafter \@secondoftwo
 \fi
}%
\providecommand \@ifx [1]{%
 \ifx #1\expandafter \@firstoftwo
 \else \expandafter \@secondoftwo
 \fi
}%
\providecommand \natexlab [1]{#1}%
\providecommand \enquote  [1]{``#1''}%
\providecommand \bibnamefont  [1]{#1}%
\providecommand \bibfnamefont [1]{#1}%
\providecommand \citenamefont [1]{#1}%
\providecommand \href@noop [0]{\@secondoftwo}%
\providecommand \href [0]{\begingroup \@sanitize@url \@href}%
\providecommand \@href[1]{\@@startlink{#1}\@@href}%
\providecommand \@@href[1]{\endgroup#1\@@endlink}%
\providecommand \@sanitize@url [0]{\catcode `\\12\catcode `\$12\catcode
  `\&12\catcode `\#12\catcode `\^12\catcode `\_12\catcode `\%12\relax}%
\providecommand \@@startlink[1]{}%
\providecommand \@@endlink[0]{}%
\providecommand \url  [0]{\begingroup\@sanitize@url \@url }%
\providecommand \@url [1]{\endgroup\@href {#1}{\urlprefix }}%
\providecommand \urlprefix  [0]{URL }%
\providecommand \Eprint [0]{\href }%
\providecommand \doibase [0]{http://dx.doi.org/}%
\providecommand \selectlanguage [0]{\@gobble}%
\providecommand \bibinfo  [0]{\@secondoftwo}%
\providecommand \bibfield  [0]{\@secondoftwo}%
\providecommand \translation [1]{[#1]}%
\providecommand \BibitemOpen [0]{}%
\providecommand \bibitemStop [0]{}%
\providecommand \bibitemNoStop [0]{.\EOS\space}%
\providecommand \EOS [0]{\spacefactor3000\relax}%
\providecommand \BibitemShut  [1]{\csname bibitem#1\endcsname}%
\let\auto@bib@innerbib\@empty
\bibitem [{\citenamefont {Sawa}(2008)}]{saw_2008}%
  \BibitemOpen
  \bibfield  {author} {\bibinfo {author} {\bibfnamefont {A.}~\bibnamefont
  {Sawa}},\ }\href {\doibase https://doi.org/10.1016/S1369-7021(08)70119-6}
  {\bibfield  {journal} {\bibinfo  {journal} {Materials Today}\ }\textbf
  {\bibinfo {volume} {11}},\ \bibinfo {pages} {28} (\bibinfo {year}
  {2008})}\BibitemShut {NoStop}%
\bibitem [{\citenamefont {Ielmini}\ and\ \citenamefont
  {Waser}(2016)}]{iel_2016}%
  \BibitemOpen
  \bibfield  {author} {\bibinfo {author} {\bibfnamefont {D.}~\bibnamefont
  {Ielmini}}\ and\ \bibinfo {author} {\bibfnamefont {R.}~\bibnamefont
  {Waser}},\ }\href@noop {} {\emph {\bibinfo {title} {Resistive Switching: From
  Fundamentals of Nanoionic Redox Processes to Memristive Device
  Applications}}}\ (\bibinfo  {publisher} {Wiley-VCH},\ \bibinfo {year}
  {2016})\BibitemShut {NoStop}%
\bibitem [{\citenamefont {Wang}\ \emph {et~al.}(2017)\citenamefont {Wang},
  \citenamefont {Joshi}, \citenamefont {Savel’ev}, \citenamefont {Jiang},
  \citenamefont {Midya}, \citenamefont {Lin}, \citenamefont {Hu}, \citenamefont
  {Ge}, \citenamefont {Strachan}, \citenamefont {Li} \emph
  {et~al.}}]{wan_2017}%
  \BibitemOpen
  \bibfield  {author} {\bibinfo {author} {\bibfnamefont {Z.}~\bibnamefont
  {Wang}}, \bibinfo {author} {\bibfnamefont {S.}~\bibnamefont {Joshi}},
  \bibinfo {author} {\bibfnamefont {S.~E.}\ \bibnamefont {Savel’ev}},
  \bibinfo {author} {\bibfnamefont {H.}~\bibnamefont {Jiang}}, \bibinfo
  {author} {\bibfnamefont {R.}~\bibnamefont {Midya}}, \bibinfo {author}
  {\bibfnamefont {P.}~\bibnamefont {Lin}}, \bibinfo {author} {\bibfnamefont
  {M.}~\bibnamefont {Hu}}, \bibinfo {author} {\bibfnamefont {N.}~\bibnamefont
  {Ge}}, \bibinfo {author} {\bibfnamefont {J.~P.}\ \bibnamefont {Strachan}},
  \bibinfo {author} {\bibfnamefont {Z.}~\bibnamefont {Li}},  \emph {et~al.},\
  }\href@noop {} {\bibfield  {journal} {\bibinfo  {journal} {Nature Mater.}\
  }\textbf {\bibinfo {volume} {16}},\ \bibinfo {pages} {101} (\bibinfo {year}
  {2017})}\BibitemShut {NoStop}%
\bibitem [{\citenamefont {Yu}(2017)}]{yu_2017}%
  \BibitemOpen
  \bibfield  {author} {\bibinfo {author} {\bibfnamefont {S.}~\bibnamefont
  {Yu}},\ }\href@noop {} {\emph {\bibinfo {title} {Neuro-inspiring computing
  using resistive synaptic devices}}}\ (\bibinfo  {publisher} {Springer
  International Publishing},\ \bibinfo {year} {2017})\BibitemShut {NoStop}%
\bibitem [{\citenamefont {Chanthbouala}\ \emph {et~al.}(2012)\citenamefont
  {Chanthbouala}, \citenamefont {Garcia}, \citenamefont {Cherifi},
  \citenamefont {Bouzehouane}, \citenamefont {Fusil}, \citenamefont {Moya},
  \citenamefont {Xavier}, \citenamefont {Yamada}, \citenamefont {Deranlot},
  \citenamefont {Mathur} \emph {et~al.}}]{chan_2012}%
  \BibitemOpen
  \bibfield  {author} {\bibinfo {author} {\bibfnamefont {A.}~\bibnamefont
  {Chanthbouala}}, \bibinfo {author} {\bibfnamefont {V.}~\bibnamefont
  {Garcia}}, \bibinfo {author} {\bibfnamefont {R.~O.}\ \bibnamefont {Cherifi}},
  \bibinfo {author} {\bibfnamefont {K.}~\bibnamefont {Bouzehouane}}, \bibinfo
  {author} {\bibfnamefont {S.}~\bibnamefont {Fusil}}, \bibinfo {author}
  {\bibfnamefont {X.}~\bibnamefont {Moya}}, \bibinfo {author} {\bibfnamefont
  {S.}~\bibnamefont {Xavier}}, \bibinfo {author} {\bibfnamefont
  {H.}~\bibnamefont {Yamada}}, \bibinfo {author} {\bibfnamefont
  {C.}~\bibnamefont {Deranlot}}, \bibinfo {author} {\bibfnamefont {N.~D.}\
  \bibnamefont {Mathur}},  \emph {et~al.},\ }\href@noop {} {\bibfield
  {journal} {\bibinfo  {journal} {Nature Mater.}\ }\textbf {\bibinfo {volume}
  {11}},\ \bibinfo {pages} {860} (\bibinfo {year} {2012})}\BibitemShut
  {NoStop}%
\bibitem [{\citenamefont {Zhuravlev}\ \emph {et~al.}(2009)\citenamefont
  {Zhuravlev}, \citenamefont {Wang}, \citenamefont {Maekawa},\ and\
  \citenamefont {Tsymbal}}]{ye_2009}%
  \BibitemOpen
  \bibfield  {author} {\bibinfo {author} {\bibfnamefont {M.~Y.}\ \bibnamefont
  {Zhuravlev}}, \bibinfo {author} {\bibfnamefont {Y.}~\bibnamefont {Wang}},
  \bibinfo {author} {\bibfnamefont {S.}~\bibnamefont {Maekawa}}, \ and\
  \bibinfo {author} {\bibfnamefont {E.~Y.}\ \bibnamefont {Tsymbal}},\
  }\href@noop {} {\bibfield  {journal} {\bibinfo  {journal} {Appl. Phys.
  Lett.}\ }\textbf {\bibinfo {volume} {95}},\ \bibinfo {pages} {052902}
  (\bibinfo {year} {2009})}\BibitemShut {NoStop}%
\bibitem [{\citenamefont {Tsymbal}\ and\ \citenamefont
  {Kohlstedt}(2006)}]{tsy_2006}%
  \BibitemOpen
  \bibfield  {author} {\bibinfo {author} {\bibfnamefont {E.~Y.}\ \bibnamefont
  {Tsymbal}}\ and\ \bibinfo {author} {\bibfnamefont {H.}~\bibnamefont
  {Kohlstedt}},\ }\href@noop {} {\bibfield  {journal} {\bibinfo  {journal}
  {Science}\ }\textbf {\bibinfo {volume} {313}},\ \bibinfo {pages} {181}
  (\bibinfo {year} {2006})}\BibitemShut {NoStop}%
\bibitem [{\citenamefont {Garcia}\ \emph {et~al.}(2009)\citenamefont {Garcia},
  \citenamefont {Fusil}, \citenamefont {Bouzehouane}, \citenamefont
  {Enouz-Vedrenne}, \citenamefont {Mathur}, \citenamefont {Barthelemy},\ and\
  \citenamefont {Bibes}}]{gar_2009}%
  \BibitemOpen
  \bibfield  {author} {\bibinfo {author} {\bibfnamefont {V.}~\bibnamefont
  {Garcia}}, \bibinfo {author} {\bibfnamefont {S.}~\bibnamefont {Fusil}},
  \bibinfo {author} {\bibfnamefont {K.}~\bibnamefont {Bouzehouane}}, \bibinfo
  {author} {\bibfnamefont {S.}~\bibnamefont {Enouz-Vedrenne}}, \bibinfo
  {author} {\bibfnamefont {N.~D.}\ \bibnamefont {Mathur}}, \bibinfo {author}
  {\bibfnamefont {A.}~\bibnamefont {Barthelemy}}, \ and\ \bibinfo {author}
  {\bibfnamefont {M.}~\bibnamefont {Bibes}},\ }\href@noop {} {\bibfield
  {journal} {\bibinfo  {journal} {Nature}\ }\textbf {\bibinfo {volume} {460}},\
  \bibinfo {pages} {81} (\bibinfo {year} {2009})}\BibitemShut {NoStop}%
\bibitem [{\citenamefont {Rouco}\ \emph {et~al.}(2020)\citenamefont {Rouco},
  \citenamefont {El~Hage}, \citenamefont {Sander}, \citenamefont {Grandal},
  \citenamefont {Seurre}, \citenamefont {Palermo}, \citenamefont {Briatico},
  \citenamefont {Collin}, \citenamefont {Trastoy}, \citenamefont {Bouzehouane}
  \emph {et~al.}}]{rou_2020}%
  \BibitemOpen
  \bibfield  {author} {\bibinfo {author} {\bibfnamefont {V.}~\bibnamefont
  {Rouco}}, \bibinfo {author} {\bibfnamefont {R.}~\bibnamefont {El~Hage}},
  \bibinfo {author} {\bibfnamefont {A.}~\bibnamefont {Sander}}, \bibinfo
  {author} {\bibfnamefont {J.}~\bibnamefont {Grandal}}, \bibinfo {author}
  {\bibfnamefont {K.}~\bibnamefont {Seurre}}, \bibinfo {author} {\bibfnamefont
  {X.}~\bibnamefont {Palermo}}, \bibinfo {author} {\bibfnamefont
  {J.}~\bibnamefont {Briatico}}, \bibinfo {author} {\bibfnamefont
  {S.}~\bibnamefont {Collin}}, \bibinfo {author} {\bibfnamefont
  {J.}~\bibnamefont {Trastoy}}, \bibinfo {author} {\bibfnamefont
  {K.}~\bibnamefont {Bouzehouane}},  \emph {et~al.},\ }\href@noop {} {\bibfield
   {journal} {\bibinfo  {journal} {Nature Commun.}\ }\textbf {\bibinfo {volume}
  {11}},\ \bibinfo {pages} {658} (\bibinfo {year} {2020})}\BibitemShut
  {NoStop}%
\bibitem [{\citenamefont {Blom}\ \emph {et~al.}(1994)\citenamefont {Blom},
  \citenamefont {Wolf}, \citenamefont {Cillessen},\ and\ \citenamefont
  {Krijn}}]{blom_1994}%
  \BibitemOpen
  \bibfield  {author} {\bibinfo {author} {\bibfnamefont {P.~W.~M.}\
  \bibnamefont {Blom}}, \bibinfo {author} {\bibfnamefont {R.~M.}\ \bibnamefont
  {Wolf}}, \bibinfo {author} {\bibfnamefont {J.~F.~M.}\ \bibnamefont
  {Cillessen}}, \ and\ \bibinfo {author} {\bibfnamefont {M.~P. C.~M.}\
  \bibnamefont {Krijn}},\ }\href@noop {} {\bibfield  {journal} {\bibinfo
  {journal} {Phys. Rev. Lett.}\ }\textbf {\bibinfo {volume} {73}},\ \bibinfo
  {pages} {2107} (\bibinfo {year} {1994})}\BibitemShut {NoStop}%
\bibitem [{\citenamefont {Meyer}\ and\ \citenamefont {Waser}(2006)}]{mey_2006}%
  \BibitemOpen
  \bibfield  {author} {\bibinfo {author} {\bibfnamefont {R.}~\bibnamefont
  {Meyer}}\ and\ \bibinfo {author} {\bibfnamefont {R.}~\bibnamefont {Waser}},\
  }\href@noop {} {\bibfield  {journal} {\bibinfo  {journal} {J. Appl. Phys.}\
  }\textbf {\bibinfo {volume} {100}},\ \bibinfo {pages} {051611} (\bibinfo
  {year} {2006})}\BibitemShut {NoStop}%
\bibitem [{\citenamefont {Pintilie}\ \emph {et~al.}(2010)\citenamefont
  {Pintilie}, \citenamefont {Stancu}, \citenamefont {Trupina},\ and\
  \citenamefont {Pintilie}}]{pin_2010}%
  \BibitemOpen
  \bibfield  {author} {\bibinfo {author} {\bibfnamefont {L.}~\bibnamefont
  {Pintilie}}, \bibinfo {author} {\bibfnamefont {V.}~\bibnamefont {Stancu}},
  \bibinfo {author} {\bibfnamefont {L.}~\bibnamefont {Trupina}}, \ and\
  \bibinfo {author} {\bibfnamefont {I.}~\bibnamefont {Pintilie}},\ }\href@noop
  {} {\bibfield  {journal} {\bibinfo  {journal} {Phys. Rev. B}\ }\textbf
  {\bibinfo {volume} {82}},\ \bibinfo {pages} {085319} (\bibinfo {year}
  {2010})}\BibitemShut {NoStop}%
\bibitem [{\citenamefont {Rault}\ \emph {et~al.}(2013)\citenamefont {Rault},
  \citenamefont {Agnus}, \citenamefont {Maroutian}, \citenamefont {Pillard},
  \citenamefont {Lecoeur}, \citenamefont {Niu}, \citenamefont {Vilquin},
  \citenamefont {Silly}, \citenamefont {Bendounan}, \citenamefont {Sirotti}
  \emph {et~al.}}]{rau_2013}%
  \BibitemOpen
  \bibfield  {author} {\bibinfo {author} {\bibfnamefont {J.~E.}\ \bibnamefont
  {Rault}}, \bibinfo {author} {\bibfnamefont {G.}~\bibnamefont {Agnus}},
  \bibinfo {author} {\bibfnamefont {T.}~\bibnamefont {Maroutian}}, \bibinfo
  {author} {\bibfnamefont {V.}~\bibnamefont {Pillard}}, \bibinfo {author}
  {\bibfnamefont {P.}~\bibnamefont {Lecoeur}}, \bibinfo {author} {\bibfnamefont
  {G.}~\bibnamefont {Niu}}, \bibinfo {author} {\bibfnamefont {B.}~\bibnamefont
  {Vilquin}}, \bibinfo {author} {\bibfnamefont {M.~G.}\ \bibnamefont {Silly}},
  \bibinfo {author} {\bibfnamefont {A.}~\bibnamefont {Bendounan}}, \bibinfo
  {author} {\bibfnamefont {F.}~\bibnamefont {Sirotti}},  \emph {et~al.},\
  }\href@noop {} {\bibfield  {journal} {\bibinfo  {journal} {Phys. Rev. B}\
  }\textbf {\bibinfo {volume} {87}},\ \bibinfo {pages} {155146} (\bibinfo
  {year} {2013})}\BibitemShut {NoStop}%
\bibitem [{\citenamefont {Hubmann}\ \emph {et~al.}(2016)\citenamefont
  {Hubmann}, \citenamefont {Li}, \citenamefont {Zhukov}, \citenamefont {von
  Seggern},\ and\ \citenamefont {Klein}}]{hub_2016}%
  \BibitemOpen
  \bibfield  {author} {\bibinfo {author} {\bibfnamefont {A.~H.}\ \bibnamefont
  {Hubmann}}, \bibinfo {author} {\bibfnamefont {S.}~\bibnamefont {Li}},
  \bibinfo {author} {\bibfnamefont {S.}~\bibnamefont {Zhukov}}, \bibinfo
  {author} {\bibfnamefont {H.}~\bibnamefont {von Seggern}}, \ and\ \bibinfo
  {author} {\bibfnamefont {A.}~\bibnamefont {Klein}},\ }\href@noop {}
  {\bibfield  {journal} {\bibinfo  {journal} {J. Physics D: Appl. Phys.}\
  }\textbf {\bibinfo {volume} {49}},\ \bibinfo {pages} {295304} (\bibinfo
  {year} {2016})}\BibitemShut {NoStop}%
\bibitem [{\citenamefont {Farokhipoor}\ and\ \citenamefont
  {Noheda}(2014)}]{far_2014}%
  \BibitemOpen
  \bibfield  {author} {\bibinfo {author} {\bibfnamefont {S.}~\bibnamefont
  {Farokhipoor}}\ and\ \bibinfo {author} {\bibfnamefont {B.}~\bibnamefont
  {Noheda}},\ }\href@noop {} {\bibfield  {journal} {\bibinfo  {journal} {APL
  Mater.}\ }\textbf {\bibinfo {volume} {2}},\ \bibinfo {pages} {056102}
  (\bibinfo {year} {2014})}\BibitemShut {NoStop}%
\bibitem [{\citenamefont {Liu}\ \emph {et~al.}(2013{\natexlab{a}})\citenamefont
  {Liu}, \citenamefont {Wang}, \citenamefont {Burton},\ and\ \citenamefont
  {Tsymbal}}]{liu_2013}%
  \BibitemOpen
  \bibfield  {author} {\bibinfo {author} {\bibfnamefont {X.}~\bibnamefont
  {Liu}}, \bibinfo {author} {\bibfnamefont {Y.}~\bibnamefont {Wang}}, \bibinfo
  {author} {\bibfnamefont {J.~D.}\ \bibnamefont {Burton}}, \ and\ \bibinfo
  {author} {\bibfnamefont {E.~Y.}\ \bibnamefont {Tsymbal}},\ }\href@noop {}
  {\bibfield  {journal} {\bibinfo  {journal} {Phys. Rev. B}\ }\textbf {\bibinfo
  {volume} {88}},\ \bibinfo {pages} {165139} (\bibinfo {year}
  {2013}{\natexlab{a}})}\BibitemShut {NoStop}%
\bibitem [{\citenamefont {Rozenberg}\ \emph {et~al.}(2010)\citenamefont
  {Rozenberg}, \citenamefont {S\'anchez}, \citenamefont {Weht}, \citenamefont
  {Acha}, \citenamefont {Gomez-Marlasca},\ and\ \citenamefont
  {Levy}}]{roz_2010}%
  \BibitemOpen
  \bibfield  {author} {\bibinfo {author} {\bibfnamefont {M.~J.}\ \bibnamefont
  {Rozenberg}}, \bibinfo {author} {\bibfnamefont {M.~J.}\ \bibnamefont
  {S\'anchez}}, \bibinfo {author} {\bibfnamefont {R.}~\bibnamefont {Weht}},
  \bibinfo {author} {\bibfnamefont {C.}~\bibnamefont {Acha}}, \bibinfo {author}
  {\bibfnamefont {F.}~\bibnamefont {Gomez-Marlasca}}, \ and\ \bibinfo {author}
  {\bibfnamefont {P.}~\bibnamefont {Levy}},\ }\href {\doibase
  10.1103/PhysRevB.81.115101} {\bibfield  {journal} {\bibinfo  {journal} {Phys.
  Rev. B}\ }\textbf {\bibinfo {volume} {81}},\ \bibinfo {pages} {115101}
  (\bibinfo {year} {2010})}\BibitemShut {NoStop}%
\bibitem [{\citenamefont {Yin}\ \emph {et~al.}(2010)\citenamefont {Yin},
  \citenamefont {Li}, \citenamefont {Liu}, \citenamefont {He}, \citenamefont
  {Zhuge}, \citenamefont {Chen}, \citenamefont {Lu}, \citenamefont {Pan},\ and\
  \citenamefont {Li}}]{yin_2010}%
  \BibitemOpen
  \bibfield  {author} {\bibinfo {author} {\bibfnamefont {K.}~\bibnamefont
  {Yin}}, \bibinfo {author} {\bibfnamefont {M.}~\bibnamefont {Li}}, \bibinfo
  {author} {\bibfnamefont {Y.}~\bibnamefont {Liu}}, \bibinfo {author}
  {\bibfnamefont {C.}~\bibnamefont {He}}, \bibinfo {author} {\bibfnamefont
  {F.}~\bibnamefont {Zhuge}}, \bibinfo {author} {\bibfnamefont
  {B.}~\bibnamefont {Chen}}, \bibinfo {author} {\bibfnamefont {W.}~\bibnamefont
  {Lu}}, \bibinfo {author} {\bibfnamefont {X.}~\bibnamefont {Pan}}, \ and\
  \bibinfo {author} {\bibfnamefont {R.-W.}\ \bibnamefont {Li}},\ }\href@noop {}
  {\bibfield  {journal} {\bibinfo  {journal} {Appl. Phys. Lett.}\ }\textbf
  {\bibinfo {volume} {97}},\ \bibinfo {pages} {042101} (\bibinfo {year}
  {2010})}\BibitemShut {NoStop}%
\bibitem [{\citenamefont {Rubi}\ \emph {et~al.}(2012)\citenamefont {Rubi},
  \citenamefont {Gomez-Marlasca}, \citenamefont {Bonville}, \citenamefont
  {Colson},\ and\ \citenamefont {Levy}}]{rub_2012}%
  \BibitemOpen
  \bibfield  {author} {\bibinfo {author} {\bibfnamefont {D.}~\bibnamefont
  {Rubi}}, \bibinfo {author} {\bibfnamefont {F.}~\bibnamefont
  {Gomez-Marlasca}}, \bibinfo {author} {\bibfnamefont {P.}~\bibnamefont
  {Bonville}}, \bibinfo {author} {\bibfnamefont {D.}~\bibnamefont {Colson}}, \
  and\ \bibinfo {author} {\bibfnamefont {P.}~\bibnamefont {Levy}},\ }\href@noop
  {} {\bibfield  {journal} {\bibinfo  {journal} {Phys. B}\ }\textbf {\bibinfo
  {volume} {407}},\ \bibinfo {pages} {3144} (\bibinfo {year}
  {2012})}\BibitemShut {NoStop}%
\bibitem [{\citenamefont {Tian}\ \emph {et~al.}(2019)\citenamefont {Tian},
  \citenamefont {Tan}, \citenamefont {Fan}, \citenamefont {Zheng},
  \citenamefont {Wang}, \citenamefont {Chen}, \citenamefont {Sun},
  \citenamefont {Chen}, \citenamefont {Qin}, \citenamefont {Zeng} \emph
  {et~al.}}]{tia_2019}%
  \BibitemOpen
  \bibfield  {author} {\bibinfo {author} {\bibfnamefont {J.}~\bibnamefont
  {Tian}}, \bibinfo {author} {\bibfnamefont {Z.}~\bibnamefont {Tan}}, \bibinfo
  {author} {\bibfnamefont {Z.}~\bibnamefont {Fan}}, \bibinfo {author}
  {\bibfnamefont {D.}~\bibnamefont {Zheng}}, \bibinfo {author} {\bibfnamefont
  {Y.}~\bibnamefont {Wang}}, \bibinfo {author} {\bibfnamefont {Z.}~\bibnamefont
  {Chen}}, \bibinfo {author} {\bibfnamefont {F.}~\bibnamefont {Sun}}, \bibinfo
  {author} {\bibfnamefont {D.}~\bibnamefont {Chen}}, \bibinfo {author}
  {\bibfnamefont {M.}~\bibnamefont {Qin}}, \bibinfo {author} {\bibfnamefont
  {M.}~\bibnamefont {Zeng}},  \emph {et~al.},\ }\href {\doibase
  10.1103/PhysRevApplied.11.024058} {\bibfield  {journal} {\bibinfo  {journal}
  {Phys. Rev. Appl.}\ }\textbf {\bibinfo {volume} {11}},\ \bibinfo {pages}
  {024058} (\bibinfo {year} {2019})}\BibitemShut {NoStop}%
\bibitem [{\citenamefont {Lu}\ \emph {et~al.}(2017)\citenamefont {Lu},
  \citenamefont {Li}, \citenamefont {Zheng}, \citenamefont {Xiao},
  \citenamefont {Lin}, \citenamefont {Li}, \citenamefont {Wang}, \citenamefont
  {Huang}, \citenamefont {Zeng}, \citenamefont {Han} \emph {et~al.}}]{lu_2017}%
  \BibitemOpen
  \bibfield  {author} {\bibinfo {author} {\bibfnamefont {W.}~\bibnamefont
  {Lu}}, \bibinfo {author} {\bibfnamefont {C.}~\bibnamefont {Li}}, \bibinfo
  {author} {\bibfnamefont {L.}~\bibnamefont {Zheng}}, \bibinfo {author}
  {\bibfnamefont {J.}~\bibnamefont {Xiao}}, \bibinfo {author} {\bibfnamefont
  {W.}~\bibnamefont {Lin}}, \bibinfo {author} {\bibfnamefont {Q.}~\bibnamefont
  {Li}}, \bibinfo {author} {\bibfnamefont {X.~R.}\ \bibnamefont {Wang}},
  \bibinfo {author} {\bibfnamefont {Z.}~\bibnamefont {Huang}}, \bibinfo
  {author} {\bibfnamefont {S.}~\bibnamefont {Zeng}}, \bibinfo {author}
  {\bibfnamefont {K.}~\bibnamefont {Han}},  \emph {et~al.},\ }\href@noop {}
  {\bibfield  {journal} {\bibinfo  {journal} {Adv. Mater}\ }\textbf {\bibinfo
  {volume} {29}},\ \bibinfo {pages} {1606165} (\bibinfo {year}
  {2017})}\BibitemShut {NoStop}%
\bibitem [{\citenamefont {Mao}\ \emph {et~al.}(2015)\citenamefont {Mao},
  \citenamefont {Song}, \citenamefont {Xiao}, \citenamefont {Gao},
  \citenamefont {Cui}, \citenamefont {Peng}, \citenamefont {Li},\ and\
  \citenamefont {Pan}}]{mao_2015}%
  \BibitemOpen
  \bibfield  {author} {\bibinfo {author} {\bibfnamefont {H.~J.}\ \bibnamefont
  {Mao}}, \bibinfo {author} {\bibfnamefont {C.}~\bibnamefont {Song}}, \bibinfo
  {author} {\bibfnamefont {L.~R.}\ \bibnamefont {Xiao}}, \bibinfo {author}
  {\bibfnamefont {S.}~\bibnamefont {Gao}}, \bibinfo {author} {\bibfnamefont
  {B.}~\bibnamefont {Cui}}, \bibinfo {author} {\bibfnamefont {J.~J.}\
  \bibnamefont {Peng}}, \bibinfo {author} {\bibfnamefont {F.}~\bibnamefont
  {Li}}, \ and\ \bibinfo {author} {\bibfnamefont {F.}~\bibnamefont {Pan}},\
  }\href {\doibase 10.1039/C5CP00421G} {\bibfield  {journal} {\bibinfo
  {journal} {Phys. Chem. Chem. Phys.}\ }\textbf {\bibinfo {volume} {17}},\
  \bibinfo {pages} {10146} (\bibinfo {year} {2015})}\BibitemShut {NoStop}%
\bibitem [{\citenamefont {Qian}\ \emph {et~al.}(2019)\citenamefont {Qian},
  \citenamefont {Fina}, \citenamefont {Sulzbach}, \citenamefont {Sánchez},\
  and\ \citenamefont {Fontcuberta}}]{qia_2019}%
  \BibitemOpen
  \bibfield  {author} {\bibinfo {author} {\bibfnamefont {M.}~\bibnamefont
  {Qian}}, \bibinfo {author} {\bibfnamefont {I.}~\bibnamefont {Fina}}, \bibinfo
  {author} {\bibfnamefont {M.~C.}\ \bibnamefont {Sulzbach}}, \bibinfo {author}
  {\bibfnamefont {F.}~\bibnamefont {Sánchez}}, \ and\ \bibinfo {author}
  {\bibfnamefont {J.}~\bibnamefont {Fontcuberta}},\ }\href@noop {} {\bibfield
  {journal} {\bibinfo  {journal} {Adv. Electr. Mater.}\ }\textbf {\bibinfo
  {volume} {5}},\ \bibinfo {pages} {1800646} (\bibinfo {year}
  {2019})}\BibitemShut {NoStop}%
\bibitem [{\citenamefont {Sulzbach}\ \emph {et~al.}(2019)\citenamefont
  {Sulzbach}, \citenamefont {Estand{\'\i}a}, \citenamefont {Long},
  \citenamefont {Lyu}, \citenamefont {Dix}, \citenamefont {G{\`a}zquez},
  \citenamefont {Chisholm}, \citenamefont {S{\'a}nchez}, \citenamefont {Fina},\
  and\ \citenamefont {Fontcuberta}}]{sul_2019}%
  \BibitemOpen
  \bibfield  {author} {\bibinfo {author} {\bibfnamefont {M.~C.}\ \bibnamefont
  {Sulzbach}}, \bibinfo {author} {\bibfnamefont {S.}~\bibnamefont
  {Estand{\'\i}a}}, \bibinfo {author} {\bibfnamefont {X.}~\bibnamefont {Long}},
  \bibinfo {author} {\bibfnamefont {J.}~\bibnamefont {Lyu}}, \bibinfo {author}
  {\bibfnamefont {N.}~\bibnamefont {Dix}}, \bibinfo {author} {\bibfnamefont
  {J.}~\bibnamefont {G{\`a}zquez}}, \bibinfo {author} {\bibfnamefont {M.~F.}\
  \bibnamefont {Chisholm}}, \bibinfo {author} {\bibfnamefont {F.}~\bibnamefont
  {S{\'a}nchez}}, \bibinfo {author} {\bibfnamefont {I.}~\bibnamefont {Fina}}, \
  and\ \bibinfo {author} {\bibfnamefont {J.}~\bibnamefont {Fontcuberta}},\
  }\href@noop {} {\bibfield  {journal} {\bibinfo  {journal} {Adv. Electr.
  Mater.}\ ,\ \bibinfo {pages} {1900852}} (\bibinfo {year} {2019})}\BibitemShut
  {NoStop}%
\bibitem [{\citenamefont {Li}\ \emph {et~al.}(2015)\citenamefont {Li},
  \citenamefont {Zhou}, \citenamefont {Jing}, \citenamefont {Zeng},
  \citenamefont {Wu}, \citenamefont {Gao}, \citenamefont {Zhang}, \citenamefont
  {Gao}, \citenamefont {Lu}, \citenamefont {Liu} \emph {et~al.}}]{li_2015}%
  \BibitemOpen
  \bibfield  {author} {\bibinfo {author} {\bibfnamefont {M.}~\bibnamefont
  {Li}}, \bibinfo {author} {\bibfnamefont {J.}~\bibnamefont {Zhou}}, \bibinfo
  {author} {\bibfnamefont {X.}~\bibnamefont {Jing}}, \bibinfo {author}
  {\bibfnamefont {M.}~\bibnamefont {Zeng}}, \bibinfo {author} {\bibfnamefont
  {S.}~\bibnamefont {Wu}}, \bibinfo {author} {\bibfnamefont {J.}~\bibnamefont
  {Gao}}, \bibinfo {author} {\bibfnamefont {Z.}~\bibnamefont {Zhang}}, \bibinfo
  {author} {\bibfnamefont {X.}~\bibnamefont {Gao}}, \bibinfo {author}
  {\bibfnamefont {X.}~\bibnamefont {Lu}}, \bibinfo {author} {\bibfnamefont
  {J.-M.}\ \bibnamefont {Liu}},  \emph {et~al.},\ }\href@noop {} {\bibfield
  {journal} {\bibinfo  {journal} {Adv. Electr. Mater.}\ }\textbf {\bibinfo
  {volume} {1}},\ \bibinfo {pages} {1500069} (\bibinfo {year}
  {2015})}\BibitemShut {NoStop}%
\bibitem [{\citenamefont {Everhardt}\ \emph {et~al.}(2016)\citenamefont
  {Everhardt}, \citenamefont {Matzen}, \citenamefont {Domingo}, \citenamefont
  {Catalan},\ and\ \citenamefont {Noheda}}]{ever_2016}%
  \BibitemOpen
  \bibfield  {author} {\bibinfo {author} {\bibfnamefont {A.}~\bibnamefont
  {Everhardt}}, \bibinfo {author} {\bibfnamefont {S.}~\bibnamefont {Matzen}},
  \bibinfo {author} {\bibfnamefont {N.}~\bibnamefont {Domingo}}, \bibinfo
  {author} {\bibfnamefont {G.}~\bibnamefont {Catalan}}, \ and\ \bibinfo
  {author} {\bibfnamefont {B.}~\bibnamefont {Noheda}},\ }\href@noop {}
  {\bibfield  {journal} {\bibinfo  {journal} {Adv. Elect. Mater.}\ }\textbf
  {\bibinfo {volume} {2}},\ \bibinfo {pages} {1500214} (\bibinfo {year}
  {2016})}\BibitemShut {NoStop}%
\bibitem [{\citenamefont {Everhardt}\ \emph {et~al.}(2019)\citenamefont
  {Everhardt}, \citenamefont {Damerio}, \citenamefont {Zorn}, \citenamefont
  {Zhou}, \citenamefont {Domingo}, \citenamefont {Catalan}, \citenamefont
  {Salje}, \citenamefont {Chen},\ and\ \citenamefont {Noheda}}]{ever_2019}%
  \BibitemOpen
  \bibfield  {author} {\bibinfo {author} {\bibfnamefont {A.~S.}\ \bibnamefont
  {Everhardt}}, \bibinfo {author} {\bibfnamefont {S.}~\bibnamefont {Damerio}},
  \bibinfo {author} {\bibfnamefont {J.~A.}\ \bibnamefont {Zorn}}, \bibinfo
  {author} {\bibfnamefont {S.}~\bibnamefont {Zhou}}, \bibinfo {author}
  {\bibfnamefont {N.}~\bibnamefont {Domingo}}, \bibinfo {author} {\bibfnamefont
  {G.}~\bibnamefont {Catalan}}, \bibinfo {author} {\bibfnamefont {E.~K.}\
  \bibnamefont {Salje}}, \bibinfo {author} {\bibfnamefont {L.-Q.}\ \bibnamefont
  {Chen}}, \ and\ \bibinfo {author} {\bibfnamefont {B.}~\bibnamefont
  {Noheda}},\ }\href@noop {} {\bibfield  {journal} {\bibinfo  {journal} {Phys.
  Rev. Lett.}\ }\textbf {\bibinfo {volume} {123}},\ \bibinfo {pages} {087603}
  (\bibinfo {year} {2019})}\BibitemShut {NoStop}%
\bibitem [{\citenamefont {Everhardt}\ \emph {et~al.}(2020)\citenamefont
  {Everhardt}, \citenamefont {Denneulin}, \citenamefont {Grünebohm},
  \citenamefont {Shao}, \citenamefont {Ondrejkovic}, \citenamefont {Zhou},
  \citenamefont {Domingo}, \citenamefont {Catalan}, \citenamefont {Hlinka},
  \citenamefont {Zuo} \emph {et~al.}}]{ever_20}%
  \BibitemOpen
  \bibfield  {author} {\bibinfo {author} {\bibfnamefont {A.~S.}\ \bibnamefont
  {Everhardt}}, \bibinfo {author} {\bibfnamefont {T.}~\bibnamefont
  {Denneulin}}, \bibinfo {author} {\bibfnamefont {A.}~\bibnamefont
  {Grünebohm}}, \bibinfo {author} {\bibfnamefont {Y.-T.}\ \bibnamefont
  {Shao}}, \bibinfo {author} {\bibfnamefont {P.}~\bibnamefont {Ondrejkovic}},
  \bibinfo {author} {\bibfnamefont {S.}~\bibnamefont {Zhou}}, \bibinfo {author}
  {\bibfnamefont {N.}~\bibnamefont {Domingo}}, \bibinfo {author} {\bibfnamefont
  {G.}~\bibnamefont {Catalan}}, \bibinfo {author} {\bibfnamefont
  {J.}~\bibnamefont {Hlinka}}, \bibinfo {author} {\bibfnamefont {J.-M.}\
  \bibnamefont {Zuo}},  \emph {et~al.},\ }\href {\doibase 10.1063/1.5122954}
  {\bibfield  {journal} {\bibinfo  {journal} {App. Phys. Rev.}\ }\textbf
  {\bibinfo {volume} {7}},\ \bibinfo {pages} {011402} (\bibinfo {year}
  {2020})}\BibitemShut {NoStop}%
\bibitem [{\citenamefont {Scott}\ \emph {et~al.}(1999)\citenamefont {Scott},
  \citenamefont {Watanabe}, \citenamefont {Hartmann},\ and\ \citenamefont
  {Lamb}}]{scott_99}%
  \BibitemOpen
  \bibfield  {author} {\bibinfo {author} {\bibfnamefont {J.~F.}\ \bibnamefont
  {Scott}}, \bibinfo {author} {\bibfnamefont {K.}~\bibnamefont {Watanabe}},
  \bibinfo {author} {\bibfnamefont {A.~J.}\ \bibnamefont {Hartmann}}, \ and\
  \bibinfo {author} {\bibfnamefont {R.~N.}\ \bibnamefont {Lamb}},\ }\href@noop
  {} {\bibfield  {journal} {\bibinfo  {journal} {Ferroelectrics}\ }\textbf
  {\bibinfo {volume} {225}},\ \bibinfo {pages} {83} (\bibinfo {year}
  {1999})}\BibitemShut {NoStop}%
\bibitem [{\citenamefont {Chen}\ \emph {et~al.}(2009)\citenamefont {Chen},
  \citenamefont {Schafranek}, \citenamefont {Wu},\ and\ \citenamefont
  {Klein}}]{chen_09}%
  \BibitemOpen
  \bibfield  {author} {\bibinfo {author} {\bibfnamefont {F.}~\bibnamefont
  {Chen}}, \bibinfo {author} {\bibfnamefont {R.}~\bibnamefont {Schafranek}},
  \bibinfo {author} {\bibfnamefont {W.}~\bibnamefont {Wu}}, \ and\ \bibinfo
  {author} {\bibfnamefont {A.}~\bibnamefont {Klein}},\ }\href@noop {}
  {\bibfield  {journal} {\bibinfo  {journal} {J. Phys. D: Appl. Phys.}\
  }\textbf {\bibinfo {volume} {42}},\ \bibinfo {pages} {215302} (\bibinfo
  {year} {2009})}\BibitemShut {NoStop}%
\bibitem [{\citenamefont {Liu}\ \emph {et~al.}(2013{\natexlab{b}})\citenamefont
  {Liu}, \citenamefont {Wang}, \citenamefont {Burton},\ and\ \citenamefont
  {Tsymbal}}]{liu_13}%
  \BibitemOpen
  \bibfield  {author} {\bibinfo {author} {\bibfnamefont {X.}~\bibnamefont
  {Liu}}, \bibinfo {author} {\bibfnamefont {Y.}~\bibnamefont {Wang}}, \bibinfo
  {author} {\bibfnamefont {J.~D.}\ \bibnamefont {Burton}}, \ and\ \bibinfo
  {author} {\bibfnamefont {E.~Y.}\ \bibnamefont {Tsymbal}},\ }\href {\doibase
  10.1103/PhysRevB.88.165139} {\bibfield  {journal} {\bibinfo  {journal} {Phys.
  Rev. B}\ }\textbf {\bibinfo {volume} {88}},\ \bibinfo {pages} {165139}
  (\bibinfo {year} {2013}{\natexlab{b}})}\BibitemShut {NoStop}%
\bibitem [{\citenamefont {Ferreyra}\ \emph {et~al.}(2020)\citenamefont
  {Ferreyra}, \citenamefont {S{\'{a}}nchez}, \citenamefont {Aguirre},
  \citenamefont {Acha}, \citenamefont {Bengi{\'{o}}}, \citenamefont {Lecourt},
  \citenamefont {Lüders},\ and\ \citenamefont {Rubi}}]{fer_2020}%
  \BibitemOpen
  \bibfield  {author} {\bibinfo {author} {\bibfnamefont {C.}~\bibnamefont
  {Ferreyra}}, \bibinfo {author} {\bibfnamefont {M.}~\bibnamefont
  {S{\'{a}}nchez}}, \bibinfo {author} {\bibfnamefont {M.}~\bibnamefont
  {Aguirre}}, \bibinfo {author} {\bibfnamefont {C.}~\bibnamefont {Acha}},
  \bibinfo {author} {\bibfnamefont {S.}~\bibnamefont {Bengi{\'{o}}}}, \bibinfo
  {author} {\bibfnamefont {J.}~\bibnamefont {Lecourt}}, \bibinfo {author}
  {\bibfnamefont {U.}~\bibnamefont {Lüders}}, \ and\ \bibinfo {author}
  {\bibfnamefont {D.}~\bibnamefont {Rubi}},\ }\href {\doibase
  10.1088/1361-6528/ab6476} {\bibfield  {journal} {\bibinfo  {journal}
  {Nanotechnology}\ }\textbf {\bibinfo {volume} {31}},\ \bibinfo {pages}
  {155204} (\bibinfo {year} {2020})}\BibitemShut {NoStop}%
\bibitem [{\citenamefont {Hu}\ \emph {et~al.}(2016)\citenamefont {Hu},
  \citenamefont {Wang}, \citenamefont {Yu},\ and\ \citenamefont
  {Wu}}]{hu_2016}%
  \BibitemOpen
  \bibfield  {author} {\bibinfo {author} {\bibfnamefont {W.~J.}\ \bibnamefont
  {Hu}}, \bibinfo {author} {\bibfnamefont {Z.}~\bibnamefont {Wang}}, \bibinfo
  {author} {\bibfnamefont {W.}~\bibnamefont {Yu}}, \ and\ \bibinfo {author}
  {\bibfnamefont {T.}~\bibnamefont {Wu}},\ }\href@noop {} {\bibfield  {journal}
  {\bibinfo  {journal} {Nature Commun.}\ }\textbf {\bibinfo {volume} {7}},\
  \bibinfo {pages} {1} (\bibinfo {year} {2016})}\BibitemShut {NoStop}%
\bibitem [{SI()}]{SI}%
  \BibitemOpen
  \href {\doibase See Supplemental Material for further details about the
  Pt/PZT/PT device electrical characterization at different frequencies} {\ See
  Supplemental Material for further details about the Pt/PZT/PT device
  electrical characterization at different frequencies}\BibitemShut {NoStop}%
\bibitem [{\citenamefont {Novkovski}\ and\ \citenamefont
  {Atanassova}(2007)}]{nov07}%
  \BibitemOpen
  \bibfield  {author} {\bibinfo {author} {\bibfnamefont {N.}~\bibnamefont
  {Novkovski}}\ and\ \bibinfo {author} {\bibfnamefont {E.}~\bibnamefont
  {Atanassova}},\ }\href@noop {} {\bibfield  {journal} {\bibinfo  {journal}
  {Semicond. Sci. Tech.}\ }\textbf {\bibinfo {volume} {22}},\ \bibinfo {pages}
  {533} (\bibinfo {year} {2007})}\BibitemShut {NoStop}%
\bibitem [{\citenamefont {Rozenberg}\ \emph {et~al.}(2004)\citenamefont
  {Rozenberg}, \citenamefont {Inoue},\ and\ \citenamefont
  {S\'anchez}}]{roz_04}%
  \BibitemOpen
  \bibfield  {author} {\bibinfo {author} {\bibfnamefont {M.}~\bibnamefont
  {Rozenberg}}, \bibinfo {author} {\bibfnamefont {I.}~\bibnamefont {Inoue}}, \
  and\ \bibinfo {author} {\bibfnamefont {M.}~\bibnamefont {S\'anchez}},\
  }\href@noop {} {\bibfield  {journal} {\bibinfo  {journal} {Phys. Rev. Lett.}\
  }\textbf {\bibinfo {volume} {92}},\ \bibinfo {pages} {178302} (\bibinfo
  {year} {2004})}\BibitemShut {NoStop}%
\bibitem [{\citenamefont {Rubi}\ \emph {et~al.}(2013)\citenamefont {Rubi},
  \citenamefont {Tesler}, \citenamefont {Alposta}, \citenamefont {Kalstein},
  \citenamefont {Ghenzi}, \citenamefont {Gomez-Marlasca}, \citenamefont
  {Rozenberg},\ and\ \citenamefont {Levy}}]{rub_2013}%
  \BibitemOpen
  \bibfield  {author} {\bibinfo {author} {\bibfnamefont {D.}~\bibnamefont
  {Rubi}}, \bibinfo {author} {\bibfnamefont {F.}~\bibnamefont {Tesler}},
  \bibinfo {author} {\bibfnamefont {I.}~\bibnamefont {Alposta}}, \bibinfo
  {author} {\bibfnamefont {A.}~\bibnamefont {Kalstein}}, \bibinfo {author}
  {\bibfnamefont {N.}~\bibnamefont {Ghenzi}}, \bibinfo {author} {\bibfnamefont
  {F.}~\bibnamefont {Gomez-Marlasca}}, \bibinfo {author} {\bibfnamefont
  {M.}~\bibnamefont {Rozenberg}}, \ and\ \bibinfo {author} {\bibfnamefont
  {P.}~\bibnamefont {Levy}},\ }\href@noop {} {\bibfield  {journal} {\bibinfo
  {journal} {Appl. Phys. Lett.}\ }\textbf {\bibinfo {volume} {103}},\ \bibinfo
  {pages} {163506} (\bibinfo {year} {2013})}\BibitemShut {NoStop}%
\bibitem [{\citenamefont {Ghenzi}\ \emph {et~al.}(2013)\citenamefont {Ghenzi},
  \citenamefont {S{\'{a}}nchez},\ and\ \citenamefont {Levy}}]{ghe_2013}%
  \BibitemOpen
  \bibfield  {author} {\bibinfo {author} {\bibfnamefont {N.}~\bibnamefont
  {Ghenzi}}, \bibinfo {author} {\bibfnamefont {M.~J.}\ \bibnamefont
  {S{\'{a}}nchez}}, \ and\ \bibinfo {author} {\bibfnamefont {P.}~\bibnamefont
  {Levy}},\ }\href {\doibase 10.1088/0022-3727/46/41/415101} {\bibfield
  {journal} {\bibinfo  {journal} {J. Phys. D: Appl. Phys.}\ }\textbf {\bibinfo
  {volume} {46}},\ \bibinfo {pages} {415101} (\bibinfo {year}
  {2013})}\BibitemShut {NoStop}%
\bibitem [{\citenamefont {Yang}\ \emph {et~al.}(2004)\citenamefont {Yang},
  \citenamefont {Dickey}, \citenamefont {Randall}, \citenamefont {Barber},
  \citenamefont {Pinceloup}, \citenamefont {Henderson}, \citenamefont {Hill},
  \citenamefont {Beeson},\ and\ \citenamefont {Skamser}}]{yang_04}%
  \BibitemOpen
  \bibfield  {author} {\bibinfo {author} {\bibfnamefont {G.~Y.}\ \bibnamefont
  {Yang}}, \bibinfo {author} {\bibfnamefont {E.~C.}\ \bibnamefont {Dickey}},
  \bibinfo {author} {\bibfnamefont {C.~A.}\ \bibnamefont {Randall}}, \bibinfo
  {author} {\bibfnamefont {D.~E.}\ \bibnamefont {Barber}}, \bibinfo {author}
  {\bibfnamefont {P.}~\bibnamefont {Pinceloup}}, \bibinfo {author}
  {\bibfnamefont {M.~A.}\ \bibnamefont {Henderson}}, \bibinfo {author}
  {\bibfnamefont {R.~A.}\ \bibnamefont {Hill}}, \bibinfo {author}
  {\bibfnamefont {J.~J.}\ \bibnamefont {Beeson}}, \ and\ \bibinfo {author}
  {\bibfnamefont {D.~J.}\ \bibnamefont {Skamser}},\ }\href@noop {} {\bibfield
  {journal} {\bibinfo  {journal} {J. Appl. Phys.}\ }\textbf {\bibinfo {volume}
  {96}},\ \bibinfo {pages} {7492} (\bibinfo {year} {2004})}\BibitemShut
  {NoStop}%
\bibitem [{\citenamefont {Chentir}\ \emph {et~al.}(2008)\citenamefont
  {Chentir}, \citenamefont {Bouyssou}, \citenamefont {Ventura}, \citenamefont
  {Guégan},\ and\ \citenamefont {Anceau}}]{chentir_08}%
  \BibitemOpen
  \bibfield  {author} {\bibinfo {author} {\bibfnamefont {M.~T.}\ \bibnamefont
  {Chentir}}, \bibinfo {author} {\bibfnamefont {E.}~\bibnamefont {Bouyssou}},
  \bibinfo {author} {\bibfnamefont {L.}~\bibnamefont {Ventura}}, \bibinfo
  {author} {\bibfnamefont {G.}~\bibnamefont {Guégan}}, \ and\ \bibinfo
  {author} {\bibfnamefont {C.}~\bibnamefont {Anceau}},\ }\href@noop {}
  {\bibfield  {journal} {\bibinfo  {journal} {Integrated Ferroelectrics}\
  }\textbf {\bibinfo {volume} {96}},\ \bibinfo {pages} {75} (\bibinfo {year}
  {2008})}\BibitemShut {NoStop}%
\bibitem [{\citenamefont {Sze}\ and\ \citenamefont {Ng}(2006)}]{sze}%
  \BibitemOpen
  \bibfield  {author} {\bibinfo {author} {\bibfnamefont {S.~M.}\ \bibnamefont
  {Sze}}\ and\ \bibinfo {author} {\bibfnamefont {K.~K.}\ \bibnamefont {Ng}},\
  }\href@noop {} {\emph {\bibinfo {title} {Physics of Semiconductor Devices}}}\
  (\bibinfo  {publisher} {John Wiley and Sons, Ltd},\ \bibinfo {year}
  {2006})\BibitemShut {NoStop}%
\bibitem [{\citenamefont {Stengel}\ \emph {et~al.}(2011)\citenamefont
  {Stengel}, \citenamefont {Aguado-Puente}, \citenamefont {Spaldin},\ and\
  \citenamefont {Junquera}}]{sten_2011}%
  \BibitemOpen
  \bibfield  {author} {\bibinfo {author} {\bibfnamefont {M.}~\bibnamefont
  {Stengel}}, \bibinfo {author} {\bibfnamefont {P.}~\bibnamefont
  {Aguado-Puente}}, \bibinfo {author} {\bibfnamefont {N.}~\bibnamefont
  {Spaldin}}, \ and\ \bibinfo {author} {\bibfnamefont {J.}~\bibnamefont
  {Junquera}},\ }\href@noop {} {\bibfield  {journal} {\bibinfo  {journal}
  {Phys. Rev. B}\ }\textbf {\bibinfo {volume} {83}},\ \bibinfo {pages} {235112}
  (\bibinfo {year} {2011})}\BibitemShut {NoStop}%
\bibitem [{\citenamefont {Chen}\ \emph {et~al.}(2005)\citenamefont {Chen},
  \citenamefont {Wu}, \citenamefont {Strozier},\ and\ \citenamefont
  {Ignatiev}}]{chen_2005}%
  \BibitemOpen
  \bibfield  {author} {\bibinfo {author} {\bibfnamefont {X.}~\bibnamefont
  {Chen}}, \bibinfo {author} {\bibfnamefont {N.~J.}\ \bibnamefont {Wu}},
  \bibinfo {author} {\bibfnamefont {J.}~\bibnamefont {Strozier}}, \ and\
  \bibinfo {author} {\bibfnamefont {A.}~\bibnamefont {Ignatiev}},\ }\href@noop
  {} {\bibfield  {journal} {\bibinfo  {journal} {Appl. Phys. Lett.}\ }\textbf
  {\bibinfo {volume} {87}},\ \bibinfo {pages} {233506} (\bibinfo {year}
  {2005})}\BibitemShut {NoStop}%
\bibitem [{\citenamefont {Desu}\ and\ \citenamefont {Vijay}(1995)}]{desu_1995}%
  \BibitemOpen
  \bibfield  {author} {\bibinfo {author} {\bibfnamefont {S.~B.}\ \bibnamefont
  {Desu}}\ and\ \bibinfo {author} {\bibfnamefont {D.~P.}\ \bibnamefont
  {Vijay}},\ }\href@noop {} {\bibfield  {journal} {\bibinfo  {journal} {Mater.
  Sci. Eng. B}\ }\textbf {\bibinfo {volume} {32}},\ \bibinfo {pages} {75}
  (\bibinfo {year} {1995})}\BibitemShut {NoStop}%
\bibitem [{\citenamefont {Dawber}\ \emph {et~al.}(2003)\citenamefont {Dawber},
  \citenamefont {Chandra}, \citenamefont {Litlewwod},\ and\ \citenamefont
  {Scott}}]{dawber_2003}%
  \BibitemOpen
  \bibfield  {author} {\bibinfo {author} {\bibfnamefont {M.}~\bibnamefont
  {Dawber}}, \bibinfo {author} {\bibfnamefont {P.}~\bibnamefont {Chandra}},
  \bibinfo {author} {\bibfnamefont {P.~B.}\ \bibnamefont {Litlewwod}}, \ and\
  \bibinfo {author} {\bibfnamefont {J.}~\bibnamefont {Scott}},\ }\href@noop {}
  {\bibfield  {journal} {\bibinfo  {journal} {J. Phys.: Condens. Matter}\
  }\textbf {\bibinfo {volume} {15}},\ \bibinfo {pages} {L393} (\bibinfo {year}
  {2003})}\BibitemShut {NoStop}%
\bibitem [{\citenamefont {Rom{\'{a}}n~Acevedo}\ \emph
  {et~al.}(2018)\citenamefont {Rom{\'{a}}n~Acevedo}, \citenamefont {Ferreyra},
  \citenamefont {S{\'{a}}nchez}, \citenamefont {Acha}, \citenamefont {R~Gay},\
  and\ \citenamefont {Rubi}}]{acevedo_2018}%
  \BibitemOpen
  \bibfield  {author} {\bibinfo {author} {\bibfnamefont {W.}~\bibnamefont
  {Rom{\'{a}}n~Acevedo}}, \bibinfo {author} {\bibfnamefont {C.}~\bibnamefont
  {Ferreyra}}, \bibinfo {author} {\bibfnamefont {M.~J.}\ \bibnamefont
  {S{\'{a}}nchez}}, \bibinfo {author} {\bibfnamefont {C.}~\bibnamefont {Acha}},
  \bibinfo {author} {\bibfnamefont {.}~\bibnamefont {R~Gay}}, \ and\ \bibinfo
  {author} {\bibfnamefont {D.}~\bibnamefont {Rubi}},\ }\href@noop {} {\bibfield
   {journal} {\bibinfo  {journal} {J. Phys. D: Appl. Phys.}\ }\textbf {\bibinfo
  {volume} {51}},\ \bibinfo {pages} {125304} (\bibinfo {year}
  {2018})}\BibitemShut {NoStop}%
\bibitem [{com()}]{com}%
  \BibitemOpen
  \href {\doibase The observed relaxations are unlikely dominated by capacitive
  effects; if we model the Pt/PZT/Pt device as an equivalent parallel RC
  circuit, from DC measurements we roughly estimate R (at $f$=0) as $\sim 140 M
  \Omega$ while C -measured with the LCR meter and frequency independent in the
  range 100Hz-1MHz- is $\sim$ 2nF. If after the writing pulse the capacitor
  discharges through R this should give a RC time constant of $\sim$ 0.2s, much
  shorter that the observed relaxations time-lenght} {\ The observed
  relaxations are unlikely dominated by capacitive effects; if we model the
  Pt/PZT/Pt device as an equivalent parallel RC circuit, from DC measurements
  we roughly estimate R (at $f$=0) as $\sim 140 M \Omega$ while C -measured
  with the LCR meter and frequency independent in the range 100Hz-1MHz- is
  $\sim$ 2nF. If after the writing pulse the capacitor discharges through R
  this should give a RC time constant of $\sim$ 0.2s, much shorter that the
  observed relaxations time-lenght}\BibitemShut {NoStop}%
\bibitem [{\citenamefont {Delimova}\ \emph {et~al.}(2006)\citenamefont
  {Delimova}, \citenamefont {Grekhov}, \citenamefont {Mashovets}, \citenamefont
  {Shin}, \citenamefont {Koo}, \citenamefont {Kim},\ and\ \citenamefont
  {Park}}]{deli_2006}%
  \BibitemOpen
  \bibfield  {author} {\bibinfo {author} {\bibfnamefont {L.}~\bibnamefont
  {Delimova}}, \bibinfo {author} {\bibfnamefont {I.}~\bibnamefont {Grekhov}},
  \bibinfo {author} {\bibfnamefont {D.}~\bibnamefont {Mashovets}}, \bibinfo
  {author} {\bibfnamefont {S.}~\bibnamefont {Shin}}, \bibinfo {author}
  {\bibfnamefont {J.~M.}\ \bibnamefont {Koo}}, \bibinfo {author} {\bibfnamefont
  {S.~P.}\ \bibnamefont {Kim}}, \ and\ \bibinfo {author} {\bibfnamefont
  {Y.}~\bibnamefont {Park}},\ }\href@noop {} {\bibfield  {journal} {\bibinfo
  {journal} {Phys. Solid State}\ }\textbf {\bibinfo {volume} {48}},\ \bibinfo
  {pages} {1182} (\bibinfo {year} {2006})}\BibitemShut {NoStop}%
\bibitem [{\citenamefont {Shang}\ \emph {et~al.}(2006)\citenamefont {Shang},
  \citenamefont {Wang}, \citenamefont {Chen}, \citenamefont {Dong},
  \citenamefont {Li},\ and\ \citenamefont {Zhang}}]{shang_2006}%
  \BibitemOpen
  \bibfield  {author} {\bibinfo {author} {\bibfnamefont {D.~S.}\ \bibnamefont
  {Shang}}, \bibinfo {author} {\bibfnamefont {Q.}~\bibnamefont {Wang}},
  \bibinfo {author} {\bibfnamefont {L.~D.}\ \bibnamefont {Chen}}, \bibinfo
  {author} {\bibfnamefont {R.}~\bibnamefont {Dong}}, \bibinfo {author}
  {\bibfnamefont {X.~M.}\ \bibnamefont {Li}}, \ and\ \bibinfo {author}
  {\bibfnamefont {W.~Q.}\ \bibnamefont {Zhang}},\ }\href@noop {} {\bibfield
  {journal} {\bibinfo  {journal} {Phys. Rev. B}\ }\textbf {\bibinfo {volume}
  {73}},\ \bibinfo {pages} {245427} (\bibinfo {year} {2006})}\BibitemShut
  {NoStop}%
\end{thebibliography}%

\end{document}